# Stellar engines and Dyson bubbles can be stable

Colin R. McInnes★
*James Watt School of Engineering, University of Glasgow, Glasgow G12 8QQ, UK*

**ABSTRACT**

A range of speculative space ventures envisage the use of ultra-large structures for the collection and reflection of light. Given the length-scale of such structures they cannot be considered as point masses for the calculation of gravitational and radiation pressure forces. Using a simplified model it will be demonstrated that ultra-large reflectors in static equilibrium levitating above a central star (so-called stellar engines) are always unstable if the reflector comprises a uniform disc. However, if the reflector has a non-uniform mass distribution, specifically a ring supporting a reflector, a stellar engine can in principle be passively stable. Moreover, while it can be shown that static swarms of reflectors levitating above a central star (so-called Dyson bubbles) are unstable, in principle they can become passively self-stabilizing if arranged about the star as a dense cloud. While such ventures are clearly speculative, understanding the orbital dynamics of ultra-large structures, and in particular the conditions for passive stability, can provide insights into the properties of potential technosignatures in search for extraterrestrial intelligence studies.

## 1. INTRODUCTION

Ultra-large structures have long been considered for a range of speculative space ventures such as asteroid orbit modification (C. R. McInnes 2007), climate engineering (W. Seifritz 1989; J. P. Sánchez Cuartielles & C. R. McInnes 2015), terraforming (R. Zubrin & C. P. McKay 1993; C. R. McInnes 2009), and planetary orbit modification (C. R. McInnes 2002). At even larger length-scales, reflectors have been considered as a means of modifying the motion of stars through gravitational coupling (L. M. Shkadov 1987), so-called stellar engines (V. Badescu & R. B. Cathcart 2006). In its simplest form a stellar engine can be considered as a flat reflective disc. C. R. McInnes (1991) demonstrated from first principles that the centre-of-mass of such a star-reflector system will accelerate in an inertial frame of reference since the two masses are gravitationally coupled, while the reflector experiences a force due to radiation pressure. Such structures would intercept and scatter a significant fraction of the energy output of the central star and so represent a potential technosignature in search for extra-terrestrial intelligence (SETI) studies (D. H. Forgan 2013; M. Lingam & A. Loeb 2020).

In this paper, the gravitational force and radiation pressure force acting on an ultra-large reflective disc will be determined from first principles using a simplified model of a perfectly reflecting rigid disc. If the radius of the reflector $R$, at distance $r$ from a central star, is much larger than the radius of the star $R_*$ such that $R \gg R_*$, the physical extent of the reflector should be considered when determining both the gravitational and the radiation pressure forces acting on the reflector. It will be demonstrated that the radiation pressure exerted on the reflector deviates from an inverse square law close to the central star when $r/R \sim 1$. Physically, light from an assumed point-like star impinges obliquely rather than normally at the edge of the ultra-large reflective disc. The functional form of both the gravitational and radiation pressure force will be used to investigate the stability properties of static equilibrium configurations, such as the stellar engine, and also orbiting reflectors. A strategy to engineer a passively stable stellar engine will be presented.

In contrast to the radiation pressure exerted on an ultra-large reflector, C. R. McInnes & J. C. Brown (1990) considered the radiation pressure exerted on a small disc of radius $R$ at distance $r$ from a central star, considered as a spherical luminous body whose radius $R_* \gg R$. It was demonstrated using radiative transfer that the radiation pressure exerted on the reflector deviates from an inverse square law close to the star when $r/R_* \sim 1$. Physically, this is due to light from the edge of the spherical star impinging obliquely rather than normally on the small reflective disc. The instances where $R \gg R_*$ and $R_* \gg R$ are limiting cases, both of which are investigated in this paper.

Moreover, at ultra-large length-scales, a Dyson sphere can be envisaged as a means of engineering the entire energy output of a star using an enveloping closed shell (F. J. Dyson 1960; J. T. Wright 2020). The stability properties of such structures have recently been investigated in the 3-body problem (C. R. McInnes, 2025). Similarly, a so-called Dyson swarm can be considered as a means of enveloping a star without the need for a monolithic shell. Such a swarm can be envisaged as an assembly of an extremely large number of reflectors in orbit around a central star, or an assembly of reflectors in static equilibrium with radiation pressure balancing gravity (so-called Dyson bubbles). The stability of the static Dyson bubble and orbiting Dyson swarm will both be investi-gated, and a strategy to engineer a passively stable Dyson bubble will be presented. Such structures would produce an infrared excess, which again could act as a potential technosignature for SETI studies (J. T. Wright et al. 2022).

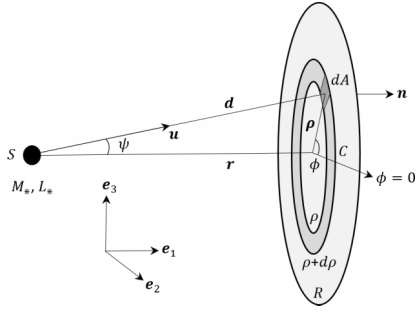

**Figure 1.** Reflecting disc of radius $R$ at distance $r$ from a body of mass $M_*$ and luminosity $L_*$.

Understanding the orbital dynamics of reflectors, and in particular the conditions required for passive stability, can provide insights into the properties of likely technosignatures (J. T. Wright et al. 2022). Passively stable equilibria and stable orbits are arguably more likely choices for such structures. Therefore, insights into the conditions required for passive stability can provide an understanding of the characteristics of potential technosignatures and help direct future searches (M. Suazo et al. 2022, 2024). Moreover, passive stability may also enable relics to survive from the past without active intervention, again providing potential technosignatures for SETI studies. The purpose of this paper is to provide insights into the underlying dynamics and stability properties of such structures using a simplified model to begin to understand how passive stability can be engineered, without exploring the details of specific astronomical examples.

The paper is organized as follows. Section 2 determines the gravitational force acting on an extended disc, while Section 3 determines the radiation pressure force acting on an extended reflective disc. Section 4 investigates the stability properties of the equilibria of a reflective disc levitating above a central star, while Section 5 considers the condition for stable circular orbits. A range of applications are discussed for stellar engines, Dyson bubbles and Dyson swarms in Section 6, with a discussion on the consequences for technosignatures and SETI in Section 7, and final conclusions drawn in Section 8.

## 2. GRAVITATIONAL FORCE ON AN EXTENDED DISC

In order to investigate the orbital dynamics of an ultra-large reflective disc, the gravitational force and radiation pressure force exerted on the disc must first be determined. As shown in Fig. 1, a uniform disc of mass $m$ and radius $R$ can be considered at a distance $r$ from a central star of mass $M_*$ and luminosity $L_*$ at position $S$, where it is assumed that $m \ll M_*$. An area element $dA = \rho d\rho d\phi$, and corresponding mass element $dm = \sigma dA$, can then be defined using polar coordinates $(\rho, \phi)$ with origin at the centre of the disc $C$, where $\sigma$ is the areal density of the disc. It is implicitly assumed that the disc is rigid, however this assumption will be discussed later in Appendix A.

The vector connecting $C$ to the mass element $dm$ is defined as $\boldsymbol{\rho} = (0, \rho \cos\phi, \rho \sin\phi)$ relative to a set of basis vectors $(\boldsymbol{e}_1, \boldsymbol{e}_2, \boldsymbol{e}_3)$. It can be seen that the mass element is located at a dis-tance $d$ from the star, where $d = \|\boldsymbol{r} + \boldsymbol{\rho}\|$. Then, the gravitational force $d\boldsymbol{f}_G$ exerted on the mass element $dm$ can be written as

$$d\boldsymbol{f}_G = -\frac{GM_*}{d^2} dm \, \boldsymbol{u}, \quad (1)$$

where $G$ is the universal gravitational constant, $d^2 = \rho^2 + r^2$ and the $\boldsymbol{u}$ is a unit vector acting along $\boldsymbol{d}$. The total gravitational force $\boldsymbol{f}_G$ exerted on the disc can then be written as

$$\boldsymbol{f}_G = -GM_* \sigma \int_0^{2\pi} \int_0^R \frac{\rho}{\rho^2 + r^2} d\rho d\phi \boldsymbol{u}. \quad (2)$$

From Fig. 1 the unit vector $\boldsymbol{u}$ can be written in component form as $(\cos\psi, \sin\psi\cos\phi, \sin\psi\sin\phi)$. Therefore, from equation (2) the last two azimuthal integrals in the vector vanish by symmetry. It can then be seen that the only non-zero component of gravitational force is directed along the line connecting $S$ and $C$. Therefore, since $\cos\psi = r/d$, it can be seen that

$$f_G = -2\pi GM_* \sigma r \int_0^R \frac{\rho}{(\rho^2 + r^2)^{3/2}} d\rho. \quad (3)$$

Then, performing the integration it can be shown that the gravitational force can be written as

$$f_G = -\frac{2GM_* m}{R^2} \left(1 - \frac{(r/R)}{\sqrt{1 + (r/R)^2}}\right), \quad (4)$$

where is the total mass of the disc $m$ is given by $\sigma \pi R^2$. Finally, defining a non-dimensional distance $\xi = r/R$ the gravitational force can now be written compactly as

$$f_G = -\bar{f}_G \left(1 - \frac{\xi}{\sqrt{1 + \xi^2}}\right), \quad (5)$$

where $\bar{f}_G = 2GM_* m/R^2$, or alternatively $\bar{f}_G = 2\pi\sigma GM_*$. It can be seen that as $r \to 0$, and so $\xi \to 0$, that $f_G \to -\bar{f}_G$. Furthermore, it can also be seen that as $R \to \infty$, and so $\xi \to 0$, that $f_G \to -\bar{f}_G$. Physically, $\bar{f}_G$ therefore represents the force due to the gravitational interaction of the central star $M_*$ with an infinite uniform plate. In the limit as $r \to \infty$ a change of variable can be used where $u = 1/\xi$, such that equation (5) can be written as $-\bar{f}_G(1 - 1/\sqrt{u^2 + 1}) \approx (\bar{f}_G/2)u^2$ since $u \to 0$. Therefore, in the limit as $r \to \infty$ it can be seen that $f_G \to -GM_* m/r^2$ and an inverse square gravitational force is recovered, as expected. It can also be noted that since $\xi = r/R$ and the half-angle $\alpha$ subtended by the disc at the central star is defined by $\tan\alpha = R/r$, that $\xi = 1/\tan\alpha$ and so $f_G = -\bar{f}_G(1 - \cos\alpha)$. Therefore $f_G \to 0$ as $\alpha \to 0$ (and $\xi \to \infty$) and $f_G \to -\bar{f}_G$ as $\alpha \to \pi/2$ (and $\xi \to 0$), as expected. The half angle subtended by the disc will be used later in Section 6.2.

In order to understand the deviation of the gravitational force from an inverse square law, equation (5) can also be written as $f_G = \hat{f}_G G(\xi)$. Here, $\hat{f}_G = -GM_* m/r^2$ is the inverse square gravitational force, which can be written as $\hat{f}_G = -\bar{f}_G/2\xi^2$. Therefore, the function $G(\xi)$, which represents the deviation from an inverse square law, can be written as

$$G(\xi) = 2\xi^2 \left(1 - \frac{\xi}{\sqrt{1 + \xi^2}}\right), \quad (6)$$

where $G(\xi) \to 1$ as $r \to \infty$ as expected, and $G(\xi) \to 0$ as $r \to 0$. The functional form of $f_G$, $\hat{f}_G$, and $G(\xi)$ are shown in Fig. 2. While the deviation from an inverse square law is only apparent for $\xi \sim$ 1, the functional form of $G(\xi)$ will have important implications for the stability of equilibria and orbits.

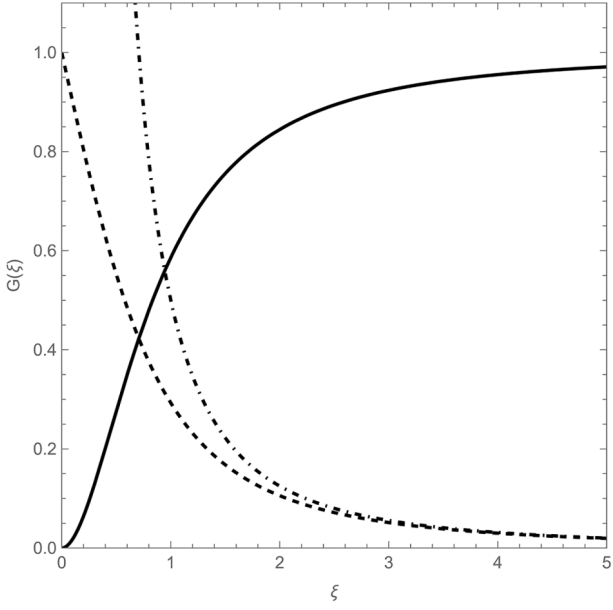

**Figure 2.** Functional form of $G(\xi)$ (———), $f_G$ (- - - -), and $\hat{f}_G$ (- · - · -) with $\bar{f}_G = 1$.

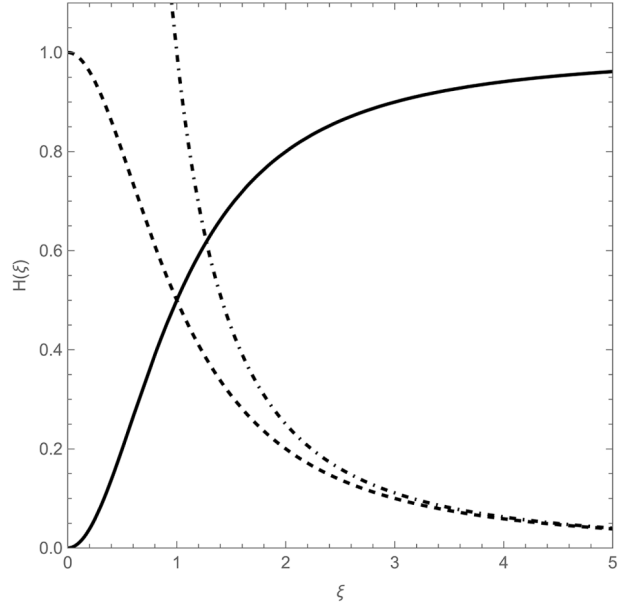

**Figure 3.** Functional form of $H(\xi)$ (———), $f_R$ (- - - -), and $\hat{f}_R$ (- · - · -) with $\bar{f}_R = 1$.

## 3. RADIATION PRESSURE FORCE ON AN EXTENDED DISC

Now that the gravitational force exerted on a disc has been determined, the radiation pressure force exerted on the disc can also be found. For ease of illustration, it will be assumed that the disc is perfectly reflecting. Again, it is implicitly assumed that the disc is rigid, however this assumption will be discussed later in Appendix A. As shown in Fig. 1, an area element $dA = \rho d\rho d\phi$ can be defined using polar coordinates $(\rho, \phi)$. Then, the unit normal to the area element is defined by $\boldsymbol{n} = (1, 0, 0)$, again relative to the unit basis vectors $(\boldsymbol{e}_1, \boldsymbol{e}_2, \boldsymbol{e}_3)$. The area element is again at a distance $d$ from a central star of luminosity $L_*$, so that the radiation pressure force $d\boldsymbol{f}_R$ exerted on the area element $dA$ can be written as (C. R. McInnes, 2004)

$$d\boldsymbol{f}_R = \frac{2}{c} \frac{L_*}{4\pi d^2} (\boldsymbol{n}.\boldsymbol{u})^2 dA \, \boldsymbol{n}, \quad (7)$$

where $c$ is the speed of light and again $\boldsymbol{u}$ is a unit vector acting along $\boldsymbol{d}$. Since the disc is assumed to be perfectly reflecting, the force $d\boldsymbol{f}_R$ is directed normal to the area element in direction $\boldsymbol{n}$. The total radiation pressure force $\boldsymbol{f}_R$ exerted on the disc can therefore be written as

$$\boldsymbol{f}_R = \frac{2}{c} \frac{L_*}{4\pi} r^2 \int_0^{2\pi} \int_0^R \frac{\rho}{(\rho^2 + r^2)^2} d\rho d\phi \, \boldsymbol{n}. \quad (8)$$

Since $\boldsymbol{n} = (1, 0, 0)$, equation (8) only has a single component along the $\boldsymbol{e}_1$ direction. Performing the integration it can be shown that the radiation pressure force can be written as

$$f_R = \frac{L_*}{2c} \left( 1 - \frac{(r/R)^2}{1 + (r/R)^2} \right). \quad (9)$$

Moreover, again defining a non-dimensional distance $\xi = r/R$, the radiation pressure force can now be written compactly as

$$f_R = \bar{f}_R \left( 1 - \frac{\xi^2}{1 + \xi^2} \right), \quad (10)$$

where $\bar{f}_R = L_*/2c$. It can be seen that as $r \to 0$, and so $\xi \to 0$, that $f_R \to \bar{f}_R$. Furthermore, it can also be seen that as $R \to \infty$, and so $\xi \to 0$, that $f_G \to \bar{f}_R$. Physically, $\bar{f}_R$ therefore represents the radiation pressure force due to the interaction of the central star of luminosity $L_*$ with an infinite uniform reflective plate (half of the momentum flux, $L_*/2c$, is intercepted). In the limit that $r \to \infty$ a change of variable can again be used where $u = 1/\xi$, such that equation (10) can be written as $\bar{f}_R(1 - 1/1 + u^2) \approx \bar{f}_R u^2$ since $u \to 0$. Therefore, in the limit that $r \to \infty$ it can be seen that $f_R \to 2L_*A/4\pi c r^2$, where $A$ is the area of the disc defined as $\pi R^2$ and an inverse square radiation pressure force for a perfectly reflecting disc is recovered, as expected. Again, it can also be noted that since $\xi = r/R$ and the half-angle $\alpha$ subtended by the disc at the central star is defined by $\tan \alpha = R/r$, then $\xi = 1/\tan \alpha$ and so $f_R = \bar{f}_R \sin^2 \alpha$. Therefore, $f_R \to 0$ as $\alpha \to 0$ (and $\xi \to \infty$) and $f_R \to \bar{f}_R$ as $\alpha \to \pi/2$ (and $\xi \to 0$), as expected.

In order to understand the deviation of the radiation pressure force from an inverse square law, equation (10) can also be written as $f_R = \hat{f}_R H(\xi)$. Here, $\hat{f}_R = 2L_*A/4\pi c r^2$ is the inverse square solar radiation pressure force, which can be written as $\hat{f}_R = \bar{f}_R/\xi^2$. Therefore, the function $H(\xi)$, which represents the deviation from an inverse square law, can be written as

$$H(\xi) = \xi^2 \left( 1 - \frac{\xi^2}{1 + \xi^2} \right), \quad (11)$$

where $H(\xi) \to 1$ as $r \to \infty$ as expected, and $H(\xi) \to 0$ as $r \to 0$. The functional form of $f_R$, $\hat{f}_R$, and $H(\xi)$ is shown in Fig. 3. Again, while the deviation from an inverse square law is only apparent for $\xi \sim 1$, the functional form of $H(\xi)$ has important implications for the stability of equilibria and orbits.

Finally, in C. R. McInnes and J. C. Brown (1990) the radiation pressure exerted on a flat reflective disc was determined due to a star considered as a finite luminous sphere whose radius $R_* \gg R$. The radiation pressure exerted on the reflector then deviates from an inverse square law when $r/R_* \sim 1$ since light from the edge of the sphere is not incident normal to the reflector. From C. R. McInnes and J. C. Brown (1990), the resulting deviation from an inverse square law $f_R = \hat{f}_R K(\zeta)$ is defined as

$$K(\zeta) = \frac{2}{3}\zeta^2 \left(1 - \left(1 - \frac{1}{\zeta^2}\right)^{3/2}\right), \quad (12)$$

where the non-dimensional distance is now defined as $\zeta = r/R_*$. It can be shown that $K(\zeta) \to 1$ as $\zeta \to \infty$ as expected, since the star becomes point like. Similarly it can be seen that $K(\zeta) \to 2/3$ as $\zeta \to 1$, representing the maximum deviation from an inverse square law as the disc approaches the surface of the central star. A more complex relationship is found for a limb darkened star, again detailed in C. R. McInnes and J. C. Brown (1990). The functional form of equation (12) is shown in Fig. 4, noting the different physical interpretations the non-dimensional distances $\xi$ and $\zeta$. It can be noted that although the star can be considered as a finite sphere in order to determine the radiation pressure force for $R_* \gg R$, gravitationally, the star is equivalent to a point mass due to Newton's shell theorem (B. C. Reed 2022). The implications of this result will be considered later in Section 4.2.

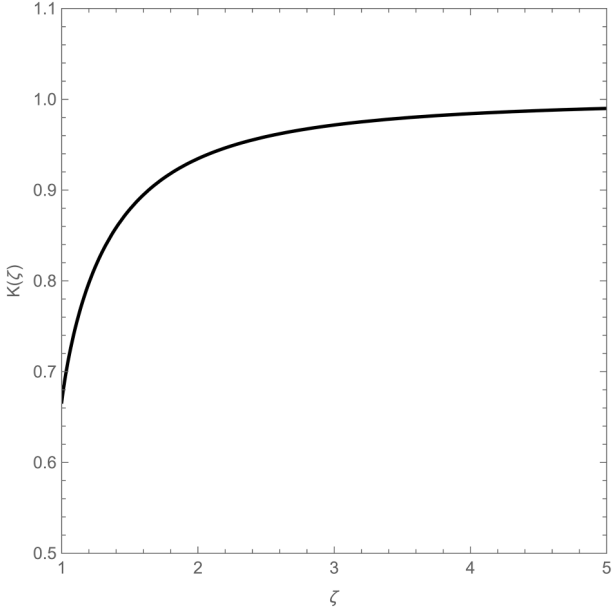

**Figure 4.** Functional form of $K(\zeta)$.

## 4. STATIC EQUILIBRIA AND STABILITY

The gravitational and radiation pressure forces exerted on an ultra-large reflective disc with $R \gg R_*$ have now been determined, and the radiation pressure force exerted on a small reflective disc with $R_* \gg R$ has been reviewed. The consequences of these two limiting cases on the stability properties of static equilibria for a reflective disc will now be investigated.

### 4.1 Large reflective disc with $R \gg R_*$

Now that the gravitational and radiation pressure forces exerted on an ideal ultra-large rigid reflective disc have been determined,

the conditions for static equilibrium can now be investigated with the reflector levitating above the central star. This will form a simple model of a stellar engine, as discussed in Section 1. Again, the assumptions of this simple model will be discussed later in Appendix A. From equations (5) and (10) a one-dimensional equation of motion can be defined such that

$$\ddot{\xi} = -\frac{\bar{f}_G}{mR}\left(1 - \frac{\xi}{\sqrt{1+\xi^2}}\right) + \frac{\bar{f}_R}{mR}\left(1 - \frac{\xi^2}{1+\xi^2}\right). \quad (13)$$

The condition for a static equilibrium configuration can then be determined simply from $\ddot{\xi} = 0$. In order to proceed, the lightness number $\beta$ of the reflector can be defined as the ratio of the radiation pressure force to the gravitational force exerted on the reflector (C. R. McInnes 2004). Since both forces in equation (13) then have an inverse square variation in the limit $\xi \to \infty$, the lightness number is defined as $\beta = \hat{f}_R/\hat{f}_G = 2\bar{f}_R/\bar{f}_G$, or alternatively $\beta = \sigma^*/\sigma$. Here, $\sigma$ is the mass per unit area (areal density) of the reflector $m/A$ and $\sigma^* = L_*/2\pi GM_*c$ is the critical areal density required for unit lightness number (C. R. McInnes 2004), again in the limit $\xi \to \infty$. It can be seen that $\sigma^*$ is only a function of the properties of the central star, while $\sigma$ is a function of the properties of the reflector. Using this definition of lightness number equation (13) can then be written as

$$\ddot{\xi} = -2\Omega^2 \left(1 - \frac{\xi}{\sqrt{1+\xi^2}}\right) + \beta\Omega^2 \left(1 - \frac{\xi^2}{1+\xi^2}\right), \quad (14)$$

where $\Omega^2 = GM_*/R^3$. The equilibria and stability properties of equation (14) can now be investigated, where the dimensional frequency $\Omega$ is retained to provide scaling for the eigenvalues of the problem.

In general, using the definition above, the lightness number required for equilibrium with $\ddot{\xi} = 0$ will be function of position as the forces deviate from an inverse square variation close to the central star. From equation (14) the lightness number $\beta = \bar{\beta}_E$ required for an equilibrium configuration at some fixed distance $\xi = \bar{\xi}$ can therefore be defined as

$$\bar{\beta}_E = 2\left(1 + \bar{\xi}^2 - \bar{\xi}\sqrt{1+\bar{\xi}^2}\right), \quad (15)$$

where $\bar{\beta}_E \to 1$ as $\bar{\xi} \to \infty$, as expected, and $\bar{\beta}_E \to 2$ as $\bar{\xi} \to 0$. Therefore, while a unit lightness number will generate the condition for equilibrium in the limit $\bar{\xi} \to \infty$, when both the gravitational and radiation pressure forces have an inverse square variation, for equilibrium close to the central star $\bar{\beta}_E \sim 2$. In general, an overscore will be used herein to denote the equilibrium value of a parameter.

Now that the condition for equilibrium has been determined, the stability properties of the equilibrium solution can be investigated. First, it will be assumed that the reflector is again in equilibrium at some position $\xi = \bar{\xi}$ where the lightness number is fixed as $\bar{\beta}_E$ using equation (15). Then, an infinitesimal perturbation can be applied such that $\xi \to \bar{\xi} + \delta\xi$ so that to linear order

$$\delta\ddot{\xi} = -2\Omega^2 \frac{\partial}{\partial \xi}\left[1 - \frac{\xi}{\sqrt{1+\xi^2}}\right]_{\xi=\bar{\xi}} \delta\xi$$
$$+ \bar{\beta}_E \Omega^2 \frac{\partial}{\partial \xi}\left[1 - \frac{\xi^2}{1+\xi^2}\right]_{\xi=\bar{\xi}} \delta\xi \quad (16)$$

which can be written as $\delta\ddot{\xi} + \Lambda^2 \delta\xi = 0$. Stability is then ensured if $\Lambda^2 > 0$ with a solution of the form $\delta\xi = \delta\bar{\xi}_1 e^{i\Lambda t} + \delta\bar{\xi}_2 e^{-i\Lambda t}$ with a pair of imaginary eigenvalues for some constants $\delta\bar{\xi}_1$ and $\delta\bar{\xi}_2$. Otherwise, the equilibrium configuration will be unstable if $\Lambda^2 < 0$ with a pair of real eigenvalues of opposite sign. From equation (16) it can be shown that

$$\Lambda^2 = 2\Omega^2 \left[ \frac{\bar{\xi}^2}{(1+\bar{\xi}^2)^{3/2}} - \frac{1}{\sqrt{1+\bar{\xi}^2}} \right]$$
$$- \bar{\beta}_E \Omega^2 \left[ \frac{2\bar{\xi}^3}{(1+\bar{\xi}^2)^2} - \frac{2\bar{\xi}}{1+\bar{\xi}^2} \right] \quad (17)$$

which then reduces to

$$\left(\frac{\Lambda}{\Omega}\right)^2 = -\frac{2}{(1+\bar{\xi}^2)^2} \left( \sqrt{1+\bar{\xi}^2} - \bar{\beta}_E \bar{\xi} \right). \quad (18)$$

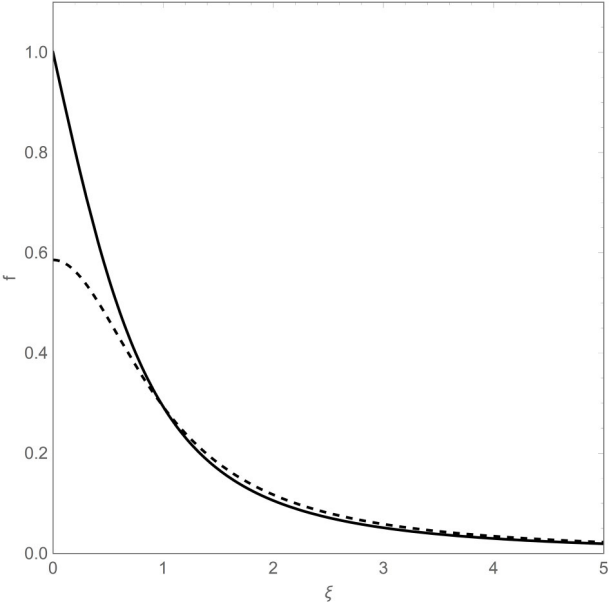

**Figure 5.** Functional form of $f_G$ (—) and $f_R$ (- - - -) for a static equilibrium configuration at $\bar{\xi} = 1$.

Using equation (15), it can be shown that the term $\sqrt{1+\bar{\xi}^2} - \bar{\beta}_E \bar{\xi}$ in equation (18) is positive if $1 + 2\bar{\xi}^2 - 2\bar{\xi}\sqrt{1+\bar{\xi}^2} > 0$. This can be shown to be true if $\bar{\xi} > 0$, so it is clear that $\Lambda^2 < 0$ and so the static equilibrium configuration is always unstable. The limiting case when $\bar{\beta}_E = 1$ occurs when $\bar{\xi} \to \infty$, where it can be shown from equation (18) that $\Lambda^2 \to 0$. This corresponds to a neutral stability configuration where the gravitational and radiation pressure forces both have an inverse square form, and so the reflector can freely translate in the radial direction. This configuration will be considered further in Section 6.3. The functional form of $f_R$ and $f_G$ are shown in Fig. 5 for a static equilibrium configuration at $\bar{\xi} = 1$, assuming $\Omega^2 = 1$ for illustration. It can be seen that for $\bar{\xi} < 1$ that $f_G > f_R$ and for $\bar{\xi} > 1$ that $f_R > f_G$, which illustrates the physical nature of the instability of the equilibrium configuration.

In order to investigate the phase space of the problem, equation (14) can be integrated to obtain the total energy of the system $E$ defined by

$$E = \frac{1}{2}\dot{\xi}^2 - 2\Omega^2 \left[\sqrt{1+\bar{\xi}^2} - \bar{\xi}\right] - \bar{\beta}_E \Omega^2 \tan^{-1}\bar{\xi}. \quad (19)$$

For a reflector in a static equilibrium configuration at $\bar{\xi} = 1$ the total energy $\bar{E} = -1.7486$, again assuming $\Omega^2 = 1$ for illustration. The resulting phase space of the problem can then be generated from equation (19), as shown in Fig 6. The fixed point is a saddle and corresponds to the unstable static equilibrium configuration, as expected. From equation (19) as $\xi \to \infty$ it can be seen that $\bar{E} \to \dot{\xi}^2/2 - \pi\bar{\beta}_E\Omega^2/2$, and so there is an asymptotic speed approximated by $\dot{\xi}_\infty = \pm\sqrt{2\bar{E} + \pi\bar{\beta}_E\Omega^2}$. For the static equilibrium configuration at $\bar{\xi} = 1$ it is found that $\dot{\xi}_\infty = \pm 0.3029$, corresponding to the stable and unstable manifolds associated with the saddle point, as can be seen in Fig. 6. While local instability has been demonstrated through linearization, the Hartman–Grobman theorem ensures that the equilibrium configuration is globally unstable, as can been seen from the structure of the phase space (A. C. King, J. Billingham & R. S. Otto 2003). Moreover, it can be noted that since the static equilibrium configuration is always unstable for displacements in the radial direction, the transverse dynamics of the reflector need not be considered to prove the instability of the problem. However, later in Section 6.2 it will be demonstrated that the static equilibrium configuration can in principle be passively stable (to linear order) if the mass distribution of the reflective disc is engineered to form a ring.

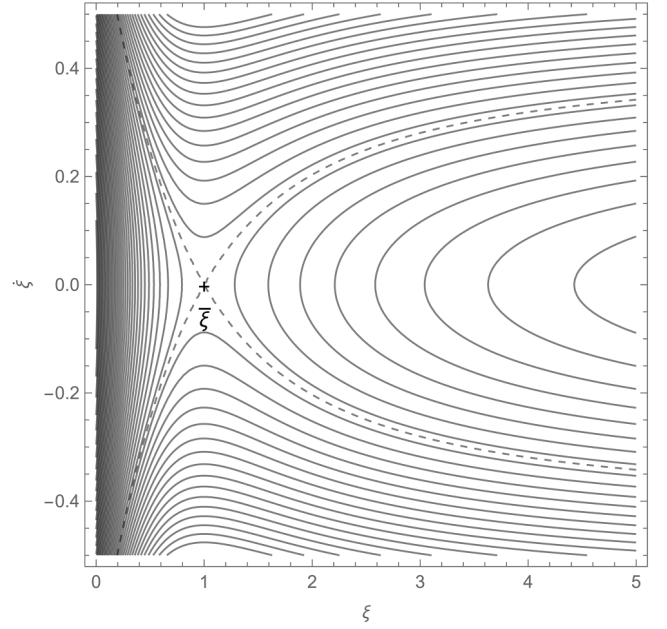

**Figure 6.** Phase space with unstable equilibrium configuration at $\bar{\xi} = 1$ and separatrix (— — —).

It can be noted that the critical reflector areal density for unit lightness number $\sigma^*$ is a function of the luminosity to mass ratio of the central star, such that $\sigma^* \propto L_*/M_*$. However, for main-sequence stars the mass–luminosity relationship scales as $L_*/L_\odot = (M_*/M_\odot)^\alpha$, where $L_\odot$ is the solar luminosity, $M_\odot$ is the solar mass, and $\alpha \sim 3.5$ (M. Harwit, 2006). Therefore, for a star of mass $M_*$ it can be shown that $\sigma^*/(\sigma^*_\odot) = (M_*/M_\odot)^{(\alpha-1)}$, where $\sigma^*_\odot$ is the critical areal density required for unit lightness number at the Sun. For example, if $M_*/M_\odot \sim 50$ then $\sigma^*/\sigma^*_\odot \sim 1.8 \times 10^4$. For the Sun, $\sigma^*_\odot \sim 1.5 \times 10^{-3}$ kg m$^{-2}$ and so for $M_*/M_\odot \sim 50$ the critical areal density is $\sigma^* \sim 27$ kg m$^{-2}$. For an aluminium reflector with a density of $2.7 \times 10^3$ kg m$^{-3}$ this corresponds to a reflector thickness of order 1 cm. Clearly, reflectors configured for massive stars would be quite different to those configured for Sun-like stars.

Finally, for an ideal reflector subject to gravitational and radiation pressure forces the gradient of these forces across the reflector will induce stresses. While the direction of the radiation pressure force is always normal to the reflector, the direction of the gravitational force will vary across the reflector moving from the centre to the edge. Therefore, while the component of the gravitational force normal to the reflector can in principle be balanced by the radiation pressure force, there will be an in-plane component of the gravitational force which will generate a compressive stress. A thin reflector will clearly be unable to support such compression. However, in principle a zero-stress reflector can be configured for a non-homogeneous, partially reflecting rotating reflector, as detailed in Appendix A.

### 4.2 Small reflective disc with $R_* \gg R$

If a small reflective disc with $R_* \gg R$ is considered then the analysis of C. R. McInnes and J. C. Brown (1990) can account for the finite size of the central star, when light from the edge of the sphere is not incident normal to the reflector. It can be noted again that although the star is considered as a finite sphere in order to determine the radiation pressure force, gravitationally it is equivalent to a point mass due to Newton's shell theorem (B. C. Reed 2022). It will be shown that the static equilibrium configuration will again be unstable with a pair of real eigenvalues. It can be shown that the one-dimensional equation of motion for the reflector can now be defined such that

$$\ddot{\zeta} = -\Gamma^2 \frac{1}{\zeta^2} + \beta \Gamma^2 \frac{1}{\zeta^2} K(\zeta), \qquad (20)$$

where $K(\zeta)$ is defined in equation (12) and where the non-dimensional distance is now $\zeta = r/R_*$ and $\Gamma^2 = GM_*/R_*^3$. The condition for static equilibrium can then be determined simply from $\ddot{\zeta} = 0$. If the reflector is in static equilibrium at position $\zeta = \bar{\zeta}$ then the required lightness number is given by $\bar{\beta}_E^* = K(\bar{\zeta})^{-1}$. Using equation (12) it can be seen that $\bar{\beta}_E^* \to 3/2$ as $\bar{\zeta} \to 1$. Following C. R. McInnes and J. C. Brown (1990), in order to determine the stability properties of the static equilibrium configuration, as before an infinitesimal perturbation can be applied such that $\zeta \to \bar{\zeta} + \delta\zeta$ so that to linear order

$$\delta\ddot{\zeta} = -2\Gamma^2 \left[\frac{1}{\zeta^3}\right]_{\zeta=\bar{\zeta}} \delta\zeta + \bar{\beta}_E^* \Gamma^2 \left[\frac{-2K(\zeta)}{\zeta^3} + \frac{1}{\zeta^2}\frac{dK(\zeta)}{d\zeta}\right]_{\zeta=\bar{\zeta}} \delta\zeta \qquad (21)$$

which can be written as $\delta\ddot{\zeta} + \Lambda^2 \delta\zeta = 0$. Again, the equilibrium configuration will be unstable if $\Lambda^2 < 0$ with a pair of real eigenvalues. For a static equilibrium configuration at position $\zeta = \bar{\zeta}$, with the required reflector lightness number $\bar{\beta}_E^* = K(\bar{\zeta})^{-1}$, it is found from equation (21) that

$$\left(\frac{\Lambda}{\Gamma}\right)^2 = -\frac{1}{\bar{\zeta}^2 K(\bar{\zeta})} \left[\frac{dK(\zeta)}{d\zeta}\right]_{\zeta=\bar{\zeta}} \qquad (22)$$

From equations (12) and (22), it is found that $\Lambda^2 < 0$ when $\bar{\zeta} > 1$ since it can be shown that $K'(\zeta) > 0$. The static equilibrium configuration is therefore unstable with a pair of real eigenvalues, as detailed in C. R. McInnes and J. C. Brown (1990). The implications of the stability properties of the static equilibrium configuration will be discussed later in Section 6.3.

## 5. CIRCULAR ORBITS AND STABILITY

Now that the stability properties of static equilibria have been investigated for the limiting cases of $R \gg R_*$ and $R_* \gg R$, the properties of circular orbits can be investigated. For the case $R \gg R_*$ both the gravitational and radiation pressure forces deviate from an inverse square law, but are still central forces, so that orbital angular momentum is conserved and the order of the problem can be reduced. For the case $R_* \gg R$ only the radiation pressure force deviates from an inverse square law. However, there are still implications for the stability of circular orbits and application such as Dyson swarms.

### 5.1 Large reflective disc with $R \gg R_*$

Following the stability analysis of Section 4.1, families of circular orbits will be now considered along with their linear stability properties. For an ideal ultra-large rigid reflective disc it can be assumed that the deviation of the radiation pressure force from an inverse square law due to the large size of the reflector dominates over the effect of the finite size of the stellar disc discussed in Section 3. From equation (14) and using plane polar coordinates $(\xi, \theta)$ the equations of motion of the problem can then be written as

$$\ddot{\xi} - \xi\dot{\theta}^2 = -2\Omega^2 \left(1 - \frac{\xi}{\sqrt{1+\xi^2}}\right) + \beta\Omega^2 \left(1 - \frac{\xi^2}{1+\xi^2}\right) \qquad (23a)$$

$$\frac{1}{\xi}\frac{d}{dt}\left(\xi^2\dot{\theta}\right) = 0 \qquad (23b)$$

where the normal to the orbiting reflector is always directed radially away from the central star, for example through passive stabilization with a slighting conical form with the centre-of-pressure displaced behind the centre-of-mass. It can be seen that while the distributed gravitational and radiation pressure forces are included in equation (23a), the centripetal and Coriolis accelerations experienced by the reflector are equivalent to that of a point mass, as noted by C. R. McInnes (2025). Moreover, it can also be noted that since equations (23) represent a central force problem, the orbital angular momentum vector will be invariant and so the orbital motion will be planar. It can be shown, but not detailed here, that the disc is stable to out-of-plane displacements to linear order.

Then, since the orbital dynamics of the reflector represents a central force problem, from equation (23b) the specific orbital angular momentum $h$ will be conserved such that $h = \xi^2\dot{\theta}$. From equation (23a) families of orbits can therefore be generated from

$$\ddot{\xi} - \frac{h^2}{\xi^3} = -2\Omega^2 \left(1 - \frac{\xi}{\sqrt{1+\xi^2}}\right) + \beta\Omega^2 \left(1 - \frac{\xi^2}{1+\xi^2}\right). \qquad (24)$$

It can be noted that the central force law represented by equation (24) is not of the type listed by F. M. Mahomed & F. Vawda (2000) which would lead directly to closed form integrability.

For a circular orbit clearly $\ddot{\xi} = 0$ and so for some fixed circular orbit of radius $\xi = \bar{\xi}$ and for some fixed lightness number $\bar{\beta}$ the orbital angular momentum $h^2$ is defined from equation (24) by

$$h^2 = 2\Omega^2\bar{\xi}^3 \left(1 - \frac{\bar{\xi}}{\sqrt{1+\bar{\xi}^2}}\right) - \bar{\beta}\Omega^2\bar{\xi}^3 \left(1 - \frac{\bar{\xi}^2}{1+\bar{\xi}^2}\right). \qquad (25)$$

For a circular orbit to exist it is clear that $h^2 > 0$. Therefore, equation (25) constraints the relationship between the circular orbit radius $\bar{\xi}$ and fixed reflector lightness number $\bar{\beta}$. Using equation (15) it can be shown from equation (25) that the constraint $h^2 > 0$ can be written as $\bar{\beta} < \bar{\beta}_E$. This is simply the reflector lightness number required for static equilibrium, and hence $h^2 = 0$.

Moreover, for a circular orbit of radius $\bar{\xi}$ with an associated circular orbit angular velocity $\omega$ it can be seen that $h = \bar{\xi}^2 \omega$, and so $\omega = h/\bar{\xi}^2$, where $h$ is defined by Eq. (25). Since $\omega$ is now a function of both the reflector orbit radius and lightness number, the reflector orbit period can be de-coupled from its orbit radius using lightness number as a free parameter. This also understood for solar sails in the two-body problem (C. R. McInnes, 2004).

Now that the condition for a circular orbit has been determined, the stability properties of the orbits can be investigated. First, it will be again be assumed that the reflector orbit radius $\xi = \bar{\xi}$, where the lightness number is selected as $\bar{\beta}$ and so the orbital angular momentum $\bar{h}$ is defined by equation (25). As before, an infinitesimal perturbation can be applied such that $\xi \to \bar{\xi} + \delta\xi$ so that to linear order

$$\delta\ddot{\xi} + \left[\frac{3\bar{h}^2}{\bar{\xi}^4}\right]_{\xi=\bar{\xi}} \delta\xi = -2\Omega^2 \frac{\partial}{\partial \xi}\left[1 - \frac{\xi}{\sqrt{1+\xi^2}}\right]_{\xi=\bar{\xi}} \delta\xi \\ + \bar{\beta}\Omega^2 \frac{\partial}{\partial \xi}\left[1 - \frac{\xi^2}{1+\xi^2}\right]_{\xi=\bar{\xi}} \delta\xi, \quad (26)$$

which can again be written as $\delta\ddot{\xi} + \Lambda^2 \delta\xi = 0$. Stability is then ensured if $\Lambda^2 > 0$ with a pair of imaginary eigenvalues. From equation (26) it can be shown that

$$\left(\frac{\Lambda}{\Omega}\right)^2 = \frac{3\bar{h}^2}{\Omega^2}\frac{1}{\bar{\xi}^4} + 2\left[\frac{\bar{\xi}^2}{(1+\bar{\xi}^2)^{3/2}} - \frac{1}{\sqrt{1+\bar{\xi}^2}}\right] \\ - \bar{\beta}\left[\frac{2\bar{\xi}^3}{(1+\bar{\xi}^2)^2} - \frac{2\bar{\xi}}{1+\bar{\xi}^2}\right], \quad (27)$$

where $\bar{h}^2$ is defined by equation (25). From equation (27) it can then be shown that the condition for a linearly stable circular orbit defined by $\Lambda^2 > 0$ can be written as $\bar{\beta} < \bar{\beta}_S$ where:

$$\bar{\beta}_S = \frac{6}{3+\bar{\xi}^2}\left[(1+2\bar{\xi}+\bar{\xi}^4) - \bar{\xi}\sqrt{1+\bar{\xi}^2}\left(\frac{4}{3}+\bar{\xi}^2\right)\right]. \quad (28)$$

In order to proceed, first it can be demonstrated that $\bar{\beta}_S < \bar{\beta}_E$. Then, for a stable, bound circular orbit it is required that $\bar{\beta} < \bar{\beta}_S$, while for a bound unstable orbit $\bar{\beta}_S < \bar{\beta} < \bar{\beta}_E$. These two regions are indicated in Fig. 7, where the band of unstable circular orbits can be seen between $\bar{\beta}_S$ and $\bar{\beta}_E$.

As an example orbit, for $\bar{\xi} = 1$ and $\bar{\beta} = 1$ it is found that $\bar{h} = 0.2929$, again assuming $\Omega^2 = 1$ for illustration. This fixed angular momentum curve is shown in the $\xi - \beta$ parameter space in Fig. 7. It can be seen that for $\bar{\beta} = 1$, there are two possible circular orbits, corresponding to the fixed point at $\bar{\xi}_1 = 1$, and also at $\bar{\xi}_2 = 2.1872$. The fixed point at $\bar{\xi}_1 = 1$ is within the stable region of the parameter space, while the fixed point at $\bar{\xi}_2 = 2.1872$ is within the unstable region of the parameter space. Moreover, for a larger specific orbital angular momentum it can be shown there will only be a single fixed point, corresponding to a single circular orbit, for a given reflector lightness number.

Again, in order to investigate the phase space of the problem, equation (24) can be integrated to determine the total energy of the system $E$ defined by

$$E = \frac{1}{2}\dot{\xi}^2 + \frac{\bar{h}^2}{2\xi^2} - 2\Omega^2\left[\sqrt{1+\xi^2} - \xi\right] + \bar{\beta}\Omega^2 \tan^{-1}\xi. \quad (29)$$

The resulting phase space of the problem can then be generated from equation (29), as shown in Fig 8 and again assuming $\Omega^2 = 1$ for illustration. For a reflector moving on the circular orbit with orbit radius $\bar{\xi}_1 = 1$ and fixed lightness number $\bar{\beta} = 1$ the total energy $\bar{E} = -1.5709$. It can be seen that the corresponding fixed point in the phase space is a stable centre. Moreover, it can seen from Fig. 8 that there is an additional fixed point at $\bar{\xi}_2 = 2.1872$, which is an unstable saddle with energy $\bar{E} = -1.5685$. These correspond to the stable and unstable orbits defined in Fig. 7. Moreover, from equation (29) it can be seen that as $\xi \to \infty$ that again $\bar{E} \to \dot{\xi}^2/2 + \pi\bar{\beta}\Omega^2/2$ and so there is an asymptotic speed approximated by $\dot{\xi}_\infty = \pm\sqrt{2\bar{E} + \pi\bar{\beta}\Omega^2}$. For the static equilibrium configuration at $\bar{\xi} = 2.1872$ and $\bar{E} = -1.5709$ and it is found that $\dot{\xi}_\infty = \pm 0.0477$, again corresponding the stable and unstable manifolds associated with the saddle point for $\xi > \bar{\xi}_2$, as can be seen in Fig. 8. For $\xi < \bar{\xi}_2$ the stable and unstable manifolds connect forming a heteroclinic connection in the phase space of the problem.

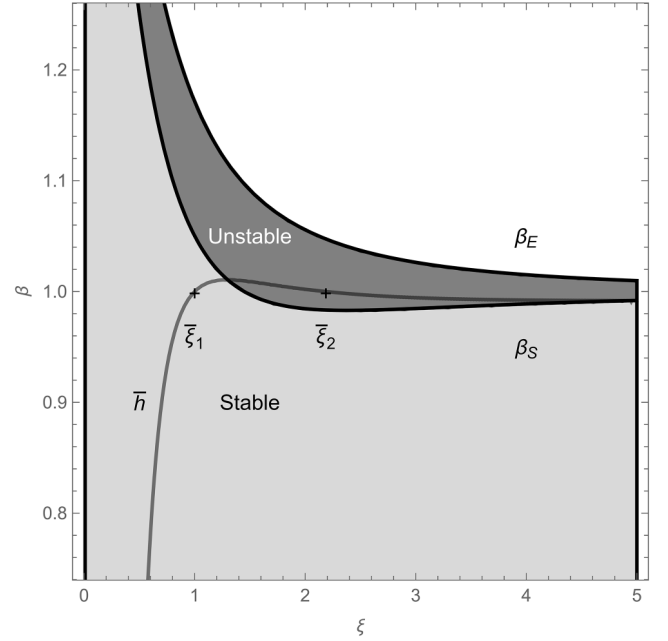

**Figure 7.** $\xi - \beta$ parameter space with stable orbits (light shaded) and unstable orbits (dark shaded).

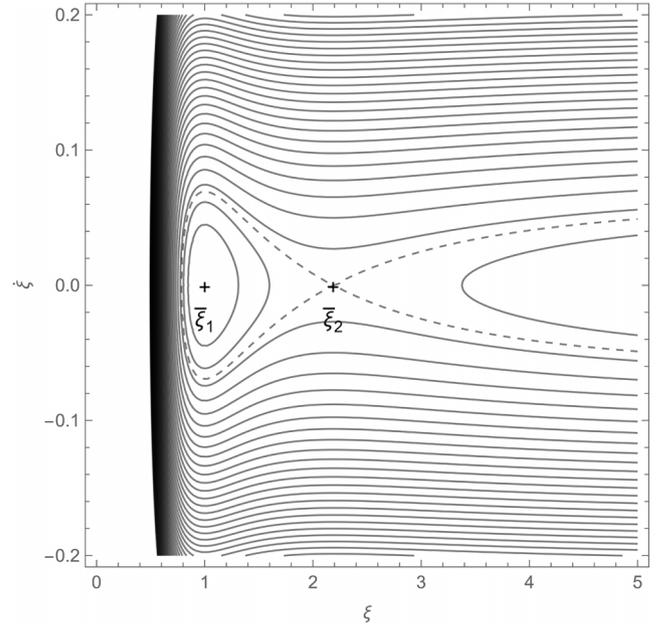

**Figure 8.** Phase space with stable circular orbit at $\bar{\xi}_1$, unstable orbit at $\bar{\xi}_2$, and separatrix $(- - -)$.

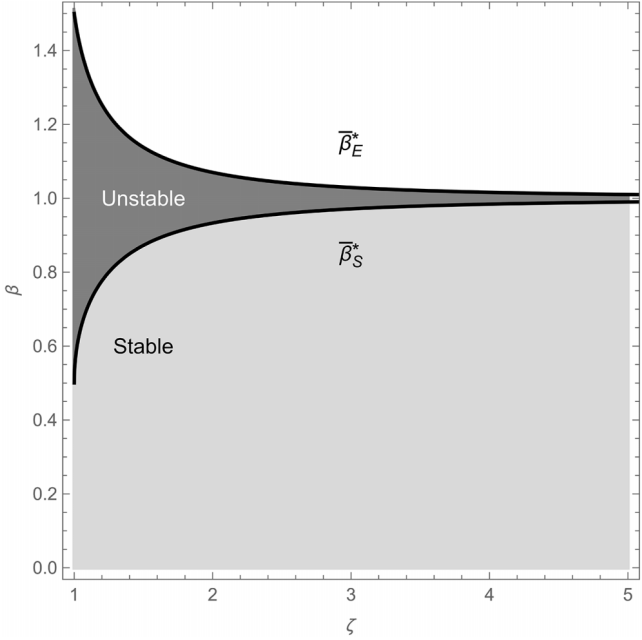

**Figure 9.** $\zeta - \beta$ parameter space with stable orbits (light shaded) and unstable orbits (dark shaded).

Now that the radial stability of the orbits have been determined, the azimuthal motion of the reflector can now be considered. If an infinitesimal perturbation $\xi \to \bar{\xi} + \delta\xi$ it applied it will lead to a modification of the orbital angular velocity $\delta\omega$ such that $\dot{\theta} \to \bar{\omega} + \delta\omega$. Then, from equation (23b) and following McInnes (1991) it can be shown that $\delta\dot{\omega} = -2\bar{\omega}\delta\dot{\xi}/\bar{\xi}$ to linear order. A reflector moving on a stable circular obit will therefore experience an azimuthal drift $\delta\theta$ defined by

$$\delta\theta(t) = \delta\theta(0) + \frac{2\bar{\omega}\delta\xi(0)}{\bar{\xi}}t - \frac{2\bar{\omega}}{\bar{\xi}}\int_0^t \delta\xi(\tau)\,\mathrm{d}\tau, \quad (30)$$

where $\delta\xi$ is the stable response defined by equation (26). The implications of the dynamics of the stable families of orbits for technosignatures will be discussed later in Section 7.3.

### 5.2 Small reflective disc with $R_* \gg R$

Now that the stability properties of circular orbits for large reflective discs have been investigated, the stability of a small reflective disc with $R_* \gg R$ can also be investigated. Again, the details of the stability analysis can be found in C. R. McInnes and J. C. Brown (1990). It is shown that the condition for a stable circular orbit for a disc of lightness number $\bar{\beta}$ in the vicinity of a star considered as a finite luminous sphere whose radius $R_* \gg R$ is given by $\bar{\beta} < \bar{\beta}_S^*$ where

$$\bar{\beta}_S^* = \left[2\bar{\zeta}^2\left(1 - \left(1 - \frac{1}{\bar{\zeta}^2}\right)^{3/2} - 2\left(1 - \frac{1}{\bar{\zeta}^2}\right)^{1/2}\right)\right]^{-1} \quad (31)$$

and again the non-dimensional distance $\bar{\zeta} = \bar{r}/R_*$ where $\bar{r}$ is the fixed circular orbit radius. The form of equation (31) is shown in Fig. 9, along with the lightness number required for static equilibrium $\bar{\beta}_E^* = K(\bar{\zeta})^{-1}$ from Section 4.2. It can be seen that there is a narrow band of instability between $\bar{\beta}_S^*$ and $\bar{\beta}_E^*$, corresponding to long period orbits close to static equilibrium. Again, the implications of the dynamics of the stable families of orbits for technosignatures will be discussed later in Section 7.3.

## 6. APPLICATIONS

The stability properties of ultra-large reflective discs with $R \gg R_*$ have been investigated, both for static equilibrium configurations and for circular orbits. Moreover, the stability properties of small reflective discs with a spherical star with $R_* \gg R$ have been reviewed, as detailed by C. R. McInnes and J. C. Brown (1990), again both for static equilibrium configurations and for circular orbits. The complete analysis for both limiting cases can therefore provide an understanding of the orbital dynamics of stellar engines and Dyson swarms and bubbles, as discussed in Section 1. The implications for the stability of such structures will now be investigated. For ease of illustration ideal reflectors will be considered, although for energy collection applications partly absorbing surfaces would be required.

### 6.1 Stellar engines

As noted in Section 1, a so-called stellar engine (L. M. Shkadov 1987) is proposed as a means of modifying the orbits of stars about the Galactic Centre. In its simplest form a stellar engine can be considered as a single ideal ultra-large rigid reflective disc in static equilibrium above a central star. Again, the assumptions of this simple model will be discussed later in Appendix A. As the disc accelerates due to radiation pressure from the star, the centre-of-mass of the gravitationally coupled star-reflector system accelerates, leading to a displacement of the star. It can be noted that the acceleration associated with the Shkadov stellar engine is extremely low since only radiation pressure is utilized. However, other stellar engine concepts propose using material from the central star as reaction mass and so generate significant larger accelerations (M. E. Caplan 2019; A. A. Svoronos 2020).

In Section 4.1, it was demonstrated that an ultra-large reflective disc in equilibrium above a central star is always unstable with $\Lambda^2 < 0$. Moreover, as $\bar{\xi} \to \infty$ then $\bar{\beta}_E \to 1$ and it can be seen from equation (18) that $\Lambda^2 \to 0$, leading to a pair of repeated zero eigenvalues. Physically, this neutral stability condition corresponds to a uniform free drift of the stellar engine since it then experiences zero net force, with both the gravitational and radiation pressure forces exhibiting inverse square laws in this limit. Moreover, for an ultra-large reflective disc it can be assumed that the derivation of the radiation pressure force from an inverse square law due to the large size of the reflector dominates over the effect of the finite size of the stellar disc discussed in Section 3. Given this inherent natural instability, active control could be envisaged to stabilize such structures. However, it is arguably of greater interest to investigate strategies for passive stabilization which do not require active interventions.

### 6.2 Passively stable stellar engines

Since it has been demonstrated that stellar engines comprising a uniform disc are always unstable, a configuration which can deliver passive stability will now be investigated. In principle, the areal density of the disc $\sigma$ can be considered as a variable such that $\sigma = \sigma(\rho)$, where $\rho$ is the radial element of the polar coordinates used to define the disc geometry in Section 2. By selecting the functional form of $\sigma$ the mass distribution of the disc can be concentrated towards the centre or edge. In the limiting cases this corresponds to a point mass at the disc centre or a uniform ring at the disc edge. These configurations are defined by $\sigma(\rho) = \bar{\sigma}\delta(\rho)$ and $\sigma(\rho) = \bar{\sigma}\delta(\rho - R)$ respectively, for some parameter $\bar{\sigma}$ where $\delta$ is the Dirac delta function. It will now be assumed that the stellar engine comprises a massive ring supporting a reflector (a stellar ring), where the mass of the reflector is assumed to be small relative to the mass of the ring. This assumption will be considered later at the end of the section. It can be noted that while the mass distribution clearly affects the gravitational force exerted on the reflector it does not affect the radiation pressure force.

The gravitational force exerted on the ring can now be found directly from equation (3), but now with a non-uniform mass distribution $\sigma(\rho) = \bar{\sigma}\delta(\rho - R)$ so that

$$f_G = -2\pi GM_* r \int_0^R \frac{\bar{\sigma}\delta(\rho - R)\rho}{(\rho^2 + r^2)^{3/2}} d\rho. \quad (32)$$

The integral can then be simplified through the Dirac delta function and so the total gravitational force exerted on the ring can be written as

$$f_G = -\frac{1}{2}\bar{f}_G \frac{\xi}{(1+\xi^2)^{3/2}} \quad (33)$$

where again $\xi = r/R$, $\bar{f}_G = 2GM_* m/R^2$ for consistency of notation with Section 2 and the total mass of the ring $m = 2\pi\bar{\sigma}R$. The radiation pressure force exerted on reflector is again defined by equation (10), therefore using the formulation of equation (14) the dynamics of the problem can be defined by

$$\ddot{\xi} = -\Omega^2 \frac{\xi}{(1+\xi^2)^{3/2}} + \beta\Omega^2\left(1 - \frac{\xi^2}{1+\xi^2}\right). \quad (34)$$

For static equilibrium the required lightness number at some equilibrium position $\xi = \bar{\xi}$ can be determined from equation (34) and can be written as $\bar{\beta}_R = \bar{\xi}/\sqrt{1+\bar{\xi}^2}$, so that $\bar{\beta}_R \to 1$ as $\bar{\xi} \to \infty$, as expected, and $\bar{\beta}_R \to 0$ as $\bar{\xi} \to 0$. In the limit that $\bar{\xi} \to 0$ the gravitational force exerted on the ring vanishes as the ring centre is superimposed on the central star.

Now that the condition for equilibrium has been determined, the stability properties of the equilibrium configuration can be investigated. First, it will again be assumed that the reflector is in equilibrium at some position $\xi = \bar{\xi}$ where the lightness number is fixed using $\bar{\beta}_R = \bar{\xi}/\sqrt{1+\bar{\xi}^2}$. Then, an infinitesimal perturbation can be applied such that $\xi \to \bar{\xi} + \delta\xi$ so that to linear order

$$\delta\ddot{\xi} = -\Omega^2 \frac{\partial}{\partial\xi}\left[\frac{\xi}{(1+\xi^2)^{3/2}}\right]_{\xi=\bar{\xi}}\delta\xi + \bar{\beta}_R\Omega^2\frac{\partial}{\partial\xi}\left[1 - \frac{\xi^2}{1+\xi^2}\right]_{\xi=\bar{\xi}}\delta\xi \quad (35)$$

which can be written as $\delta\ddot{\xi} + \Lambda^2\delta\xi = 0$. Again, stability is then ensured if $\Lambda^2 > 0$ with a pair of imaginary eigenvalues, otherwise the equilibrium configuration will be unstable if $\Lambda^2 < 0$ with a pair of real eigenvalues. From equation (35) and using $\bar{\beta}_R = \bar{\xi}/\sqrt{1+\bar{\xi}^2}$ it can be shown that

$$\left(\frac{\Lambda}{\Omega}\right)^2 = (1+\bar{\xi}^2)^{-5/2}. \quad (36)$$

Then, from equation (36), it is clear that $\Lambda^2 > 0$ and so the static equilibrium configuration is always stable. The functional form of $f_R$ and $f_G$ are shown in Fig. 10 for a static equilibrium configuration at $\bar{\xi} = 1$, again assuming $\Omega^2 = 1$ for illustration.

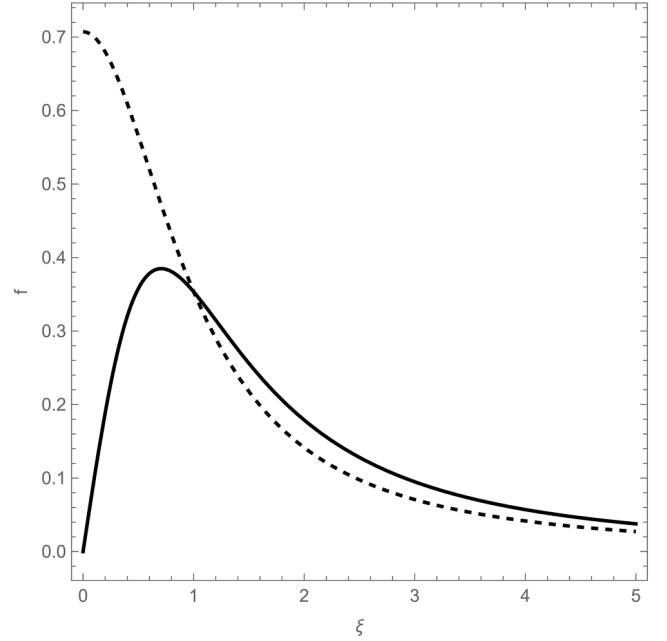

**Figure 10.** Functional form of $f_G$ (———) and $f_R$ (- - - -) for a static equilibrium configuration at $\bar{\xi} = 1$.

It can be seen that for $\bar{\xi} < 1$ that $f_R > f_G$ and for $\bar{\xi} > 1$ that $f_G > f_R$, which illustrates the physical nature of the stability of the equilibrium configuration. In principle a ring stellar engine can therefore be configured to be passively stable. In contrast, the uniform stellar engine had interaction forces shown in Fig. 5 where it can be seen that for $\bar{\xi} < 1$ that $f_R < f_G$ and for $\bar{\xi} > 1$ that $f_G < f_R$ resulting in instability. Although not detailed here, the other limiting case of a point mass at the centre of the disc with $\sigma(\rho) = \bar{\sigma}\delta(\rho)$ is found to be unstable.

Again, in order to investigate the phase space of the problem, equation (34) can be integrated to determine the total energy of the system $E$ defined by

$$E = \frac{1}{2}\dot{\xi}^2 - 2\Omega^2\frac{1}{\sqrt{1+\xi^2}} - \bar{\beta}_R\Omega^2\tan^{-1}\xi. \quad (37)$$

The resulting phase space of the problem can then be generated from equation (37), as shown in Fig. 11, where the fixed point is now a centre and corresponds to a stable equilibrium configuration with closed phase curves, again assuming $\Omega^2 = 1$ for illustration. This is in contrast to a reflector with a uniform mass distribution discussed in Section 4.1 which has an unstable saddle, as shown in Fig. 6.

Now that the radial stability of the problem has been determined, to complete the analysis the transverse dynamics of the problem must also be investigated. An infinitesimal transverse displacement $\delta\eta$ is now applied to the reflector orthogonal to $\xi$ with the reflector orientation fixed, for example through passive spin stabilization, so that the radiation pressure force will still be directed along the $\xi$-axis. The gravitational and radiation pressure forces can then be re-calculated in order to assess the transverse stability of the equilibrium configuration. From Appendix B, the radial gravitational and radiation pressure forces are found to be unchanged to linear order and so the radial forces remain in equilibrium. Since the reflector normal is fixed there is no transverse radiation pressure force. From the transverse gravitational force defined in equation (B4) in Appendix B using non-dimensional

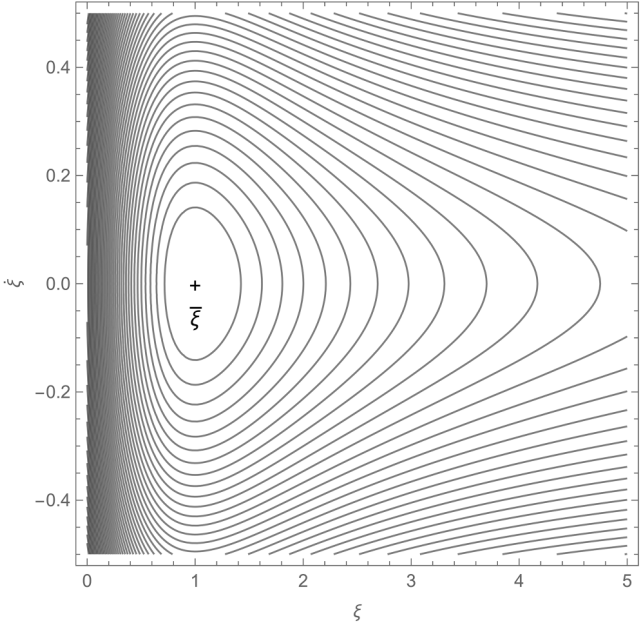

**Figure 11.** Phase space with stable equilibrium configuration for ring reflector at $\bar{\xi} = 1$.

variables it is found that

$$\delta\ddot{\eta} + \frac{1}{2}\Omega^2 \frac{2\bar{\xi}^2 - 1}{(1+\bar{\xi}^2)^{5/2}} \delta\eta = 0. \tag{38}$$

It can be seen from equation (38) that the transverse dynamics are stable if $2\bar{\xi}^2 - 1 > 0$. Since the radial dynamics are always stable, the ring stellar engine is therefore linearly stable only if $\bar{\xi} > 1/\sqrt{2}$. The transition from stability to instability can be understood from the potential of the ring. D. Schumayer and D. A. W. Hutchinson (2019) note that for displacements above a uniform ring such that $\bar{\xi} > 1/\sqrt{2}$ the minimum of the gravitational potential lies on the axis of the ring, which corresponds to the stable transverse dynamics as noted above. However, as the displacement distance above the ring falls such that $\bar{\xi} \leq 1/\sqrt{2}$, the axis of the ring becomes a potential maximum resulting in instability. This can be understood physically since for $\bar{\xi} \gg 1$ the ring is point like, and so for small transverse displacements the gravitational force is towards the central star. However, when the ring plane intersects the central star at $\bar{\xi} = 0$ the configuration is unstable, as noted by Maxwell and others (J. C. Maxwell 1859; C. R. McInnes 2025). The condition $2\bar{\xi}^2 - 1 = 0$ defines the boundary between these two states.

It has been noted that stellar engines would exhibit a distinctive technosignature (D. H. Forgan 2013; M. Lingam and A. Loeb 2020). Moreover, for a passively stable ring stellar engine the displacement distance for stability is defined by $\bar{\xi} > 1/\sqrt{2}$. As noted in Section 3, if the half-angle $\alpha$ subtended by the disc at the central star is defined by $\tan\alpha = R/r$, then $\tan\alpha = 1/\bar{\xi}$. The minimum stable configuration for the ring stellar engine therefore corresponds to a half angle of up to $\alpha \approx 55^o$. The geometry of such a passively stable stellar engine could therefore constrain the form of a likely technosignature. The implications for technosignatures will be discussed further in Section 7.1.

The assumption that the mass of the reflector is small relative to the mass of the ring can now be considered. It will be assumed that both the reflector and ring have the same bulk density $\rho_R$ and that the reflector has radius $R_R$ and thickness $h$. It will also be assumed that the ring comprises a torus of minor radius $R_T$. The mass of reflector is then $\pi R_R^2 h \rho_R$ and the mass of the torus is $2\pi R_R \pi R_T^2 \rho_R$, so that the ratio of the mass of reflector to the mass of ring is given by $\lambda = R_R h/2\pi R_T^2$. Moreover, the areal density of the entire assembly is found to be $\sigma = \rho_R(h + 2\pi R_T^2/R_R)$. However, for a minimally stable static equilibrium configuration with $\bar{\xi} = 1/\sqrt{2}$, the lightness number of the reflector is $\bar{\beta}_R = 1/\sqrt{3}$ and so the required areal density is given by $\sqrt{3}\sigma^*$. The inner radius of the torus can then be determined from $R_T = \sqrt{((k-1)hR/2\pi)}$, where $k = \sqrt{3}\sigma^*/\rho_R h$. Using this result, it can be shown that the ratio of the mass of reflector to the mass of ring is given $\lambda = 1/(k-1)$, which is independent of the radius of the reflector. While the critical areal density is a function of the luminosity and mass of the central star, for the Sun $\sigma^* = 1.53 \times 10^{-3}$ kg m$^{-2}$. If the reflector is assumed to be fabricated from aluminium with $\rho_R = 2700$ kg m$^{-3}$, then a thin film of aluminium will remain reflective at a thickness of $h = 15$ nm (K. E. Drexler 1979). It is then found that $\lambda = 0.015$, and so the assumption that the mass of the reflector is small relative to the mass of the ring appears to be reasonable. Sub-micron perforations in the reflector film, with a length-scale smaller than optical wavelengths, could also reduce the reflector mass while maintaining its reflectivity, thus further reducing the mass ratio $\lambda$. Moreover, arrays of quarter-wave pillars on the rear surface of the reflector could in principle radiate in the infrared to provide passive thermal control (C. R. McInnes 2004).

### 6.3 Dyson bubbles

As noted in Section 1, a so-called Dyson bubble can be considered as a means of enveloping a star to collect and process energy without the need for a rigid monolithic shell. The 'bubble' is assumed to comprise of a large number of small reflectors in static equilibrium about the star with radiation pressure balancing gravity. It will first be demonstrated that such structures are in general unstable, while a passively self-stabilizing bubble will be presented in the next section. It will be assumed that the central star is only enveloped by the Dyson bubble, and indeed the mass of any associated planetary system may have been used to assemble the Dyson bubble itself. Moreover, it will be assumed that the individual reflectors are of sufficiently low mass and sufficiently distant that gravitational interactions between reflectors can be ignored, although the self-gravity of a dense Dyson bubble will be considered later in Section 6.5.

First, assuming an inverse square law for both gravity $\hat{f}_G$ and radiation pressure $\hat{f}_R$, as defined in Sections 2 and 3, the dynamics of a small reflector far from the central star is defined by

$$\ddot{\bar{\xi}} = -\Omega^2 \frac{1}{\bar{\xi}^2} + \beta\Omega^2 \frac{1}{\bar{\xi}^2}, \tag{39}$$

where $\ddot{\bar{\xi}} = 0$ when a fixed lightness number $\bar{\beta} = 1$ is selected. A static equilibrium configuration will therefore be generated and, since both forces have an inverse square form, there will zero nett force at any distance $\bar{\xi}$ from the star so that $\bar{\beta}$ is independent of $\bar{\xi}$.

In order to determine the stability properties of the static equilibrium configuration, as before an infinitesimal perturbation can be applied such that $\xi \to \bar{\xi} + \delta\xi$ so that to linear order

$$\delta\ddot{\xi} = 2\Omega^2 \left[\frac{1}{\xi^3}\right]_{\xi=\bar{\xi}} \delta\xi - 2\bar{\beta}\Omega^2 \left[\frac{1}{\xi^3}\right]_{\xi=\bar{\xi}} \delta\xi \qquad (40)$$

which can again be written as $\delta\ddot{\xi} + \Lambda^2 \delta\xi = 0$, where $(\Lambda/\Omega)^2 = -2(1-\bar{\beta})/\bar{\xi}^3$. Since $\bar{\beta} = 1$ for equilibrium, independent of $\bar{\xi}$, it can be seen that $(\Lambda/\Omega)^2 = 0$ corresponding to neutral stability with repeated zero eigenvalues. This is expected since the gravitational and radiation pressure forces both have an inverse square variation. A reflector with $\bar{\beta} = 1$ will always be in equilibrium and the reflector will simply drift if some small initial radial velocity perturbation is applied.

Moreover, if the bubble is assembled from a large number of small reflectors with $R_* \gg R$ and $r/R_* \sim 1$, then the analysis of Section 4.2 can also account for the finite size of the star. As demonstrated in Section 4.2 the static equilibrium configuration is unstable with a pair of real eigenvalues. Moreover, if the bubble is assembled from a small number of large reflective discs, with $R \gg R_*$, the analysis of Section 4.1 has also demonstrated that such static equilibria will always be unstable, again with a pair of real eigenvalues (although Section 6.2 demonstrates how to passively stabilize such equilibria). Therefore, in both limiting cases, $R \gg R_*$ and $R_* \gg R$, a Dyson bubble in a static equilibrium configuration will be unstable. Again, given this natural instability, active control could be envisaged to stabilize such structures. However, it is arguably of greater interest to investigate strategies for passive stabilization which do not require interventions.

### 6.4 Self-stabilizing Dyson bubbles

It has been demonstrated that a static Dyson bubble is apparently always unstable. However, a Dyson bubble can be envisaged as comprising truly vast numbers of reflectors, which will impact the radiation pressure exerted on each reflector due to mutual attenuation. B. C. Lacki (2016) considers the deployment of a vast cloud of dipoles on a galactic scale. In order to investigate a mechanism for a self-stabilizing Dyson bubble it will therefore be assumed that a uniform, dense cloud of (small) reflectors is deployed about the central star. Moreover, it will be assumed that the cloud is of sufficient density that it attenuates the transmission of light form the central star through the cloud, but is of sufficiently low mass that gravitational interaction between the reflectors can be ignored. Again, the self-gravity of a dense Dyson bubble will be considered later in Section 6.5, as will the effect of a background of diffuse scattered radiation. For a uniform number density of reflectors $n$ and path length through the cloud $L$ the column density is $\Sigma = nL$. Therefore, if each reflector has an area A the optical depth of the cloud is given by $\tau = A\Sigma$ (G. B. Rybicki and A. P. Lightman 2004). Since the path length at distance $r$ from the central star is simply $r = \xi R$, the optical depth can be then written as $\tau = nAR\xi$. It will be assumed that cloud is homogeneous and extends inwards to the central star. While thermal limitations would inevitably limit the inner radius of such a cloud, if the outer radius of the cloud is large then it can be considered as a uniform sphere enveloping the central star.

It will now be assumed that the individual reflectors comprising the cloud are small, so that $R_* \gg R$, and that the deviation from an inverse square law for radiation pressure is dominated by the optical depth of the cloud. Therefore, for an inverse square law for both gravity $\hat{f}_G$ and radiation pressure $\hat{f}_R$, as defined in Sections 2 and 3, but with attenuation due to the optical depth of the cloud, the radiation pressure force can be written as

$$f_R = \bar{f}_R \frac{1}{\xi^2} e^{-\mu\xi}, \qquad (41)$$

where $\mu = nAR$ is now a dimensionless attenuation coefficient. Therefore, using the formulation of equation (14), the dynamics of a reflector can now be defined by

$$\ddot{\xi} = -\Omega^2 \frac{1}{\xi^2} + \beta\Omega^2 \frac{1}{\xi^2} e^{-\mu\xi}. \qquad (42)$$

For static equilibrium at $\xi = \bar{\xi}$ the required lightness number $\bar{\beta}_D$ can be determined from equation (42) and can be written as $\bar{\beta}_D = e^{\mu\bar{\xi}}$, where $\bar{\beta}_D \to \infty$ as $\bar{\xi} \to \infty$ and $\bar{\beta}_D \to 1$ as $\bar{\xi} \to 0$. In the limit that $\bar{\xi} \to \infty$ the radiation pressure from the central star vanishes due to the optical depth of the cloud, whereas in the limit that $\bar{\xi} \to 0$ the attenuation of the cloud vanishes.

Now that the condition for equilibrium has been determined, the stability properties of the equilibrium configuration can be investigated. First, it will again be assumed that the reflector is in equilibrium at some position $\xi = \bar{\xi}$ where the lightness number is fixed using $\bar{\beta}_D = e^{\mu\bar{\xi}}$. Then, an infinitesimal perturbation can be applied such that $\xi \to \bar{\xi} + \delta\xi$ so that to linear order

$$\delta\ddot{\xi} = -\Omega^2 \frac{\partial}{\partial\xi} \left[\frac{1}{\xi^2}\right]_{\xi=\bar{\xi}} \delta\xi + \bar{\beta}_D \Omega^2 \frac{\partial}{\partial\xi} \left[\frac{e^{-\mu\xi}}{\xi^2}\right]_{\xi=\bar{\xi}} \delta\xi \qquad (43)$$

which can be written as $\delta\ddot{\xi} + \Lambda^2 \delta\xi = 0$. Stability is again ensured if $\Lambda^2 > 0$ with a pair of imaginary eigenvalues, otherwise the equilibrium configuration will be unstable if $\Lambda^2 < 0$ with a pair of real eigenvalues. It can be shown that

$$\left(\frac{\Lambda}{\Omega}\right)^2 = \mu \frac{1}{\bar{\xi}^2}. \qquad (44)$$

From equation (44), it is clear that $\Lambda^2 > 0$ and so the static equilibrium configuration is always stable. In the limit $\mu \to 0$ it can be seen that $\bar{\beta}_D \to 1$ and $(\Lambda/\Omega)^2 \to 0$, which corresponds to the static equilibrium configuration with a pair of repeated zero eigenvalues, as discussed in Section 6.3. The functional form of $f_R$ and $f_G$ are shown in Fig. 12 for a static equilibrium configuration at $\bar{\xi} = 1$, assuming $\Omega^2 = 1$ and $\mu = 1$ for illustration. It can be seen that for $\bar{\xi} < 1$ that $f_R > f_G$ and for $\bar{\xi} > 1$ that $f_G > f_R$, which illustrates the physical nature of the stability of the equilibrium configuration. It can also be noted that $\Lambda^2 > 0$ for any positive $\mu$, so passive stability is in principle not dependent on a lower bound for $\mu$, although for a small number of reflectors the assumption of a uniform cloud will break down.

Again, in order to investigate the phase space of the problem, equation (42) can be integrated to determine the total energy of the system $E$ defined by

$$E = \frac{1}{2}\dot{\xi}^2 - \frac{\Omega^2}{\xi} \left[1 - \bar{\beta}_D e^{-\mu\xi}\right] + \mu\bar{\beta}\text{Ei}(-\mu\xi), \qquad (45)$$

where Ei is the exponential integral function. The resulting phase space of the problem can then be generated from equation (45), as shown in Fig. 13, where the fixed point is a centre and corresponds to the stable equilibrium configuration with closed phase curves, again assuming $\Omega^2 = 1$ and $\mu = 1$ for illustration.

Now that the radial stability of the problem has been determined, to complete the analysis the transverse dynamics must also be considered. An infinitesimal transverse displacement $\delta\eta$ is now applied to the reflector orthogonal to $\xi$ with the reflector orientation fixed, for example through passive spin stabilization, so that the radiation pressure force will still be directed along the $\xi$-axis. The radial gravitational and radiation pressure forces are again found to be unchanged to linear order and so the radial forces remain in equilibrium. Since the reflector normal is fixed there is no transverse radiation pressure force. From the transverse gravitational force defined in equation (C3b) in Appendix C using non-dimensional variables it is found that

$$\delta\ddot{\eta} + \frac{\Omega^2}{\bar{\xi}^3}\delta\eta = 0. \tag{46}$$

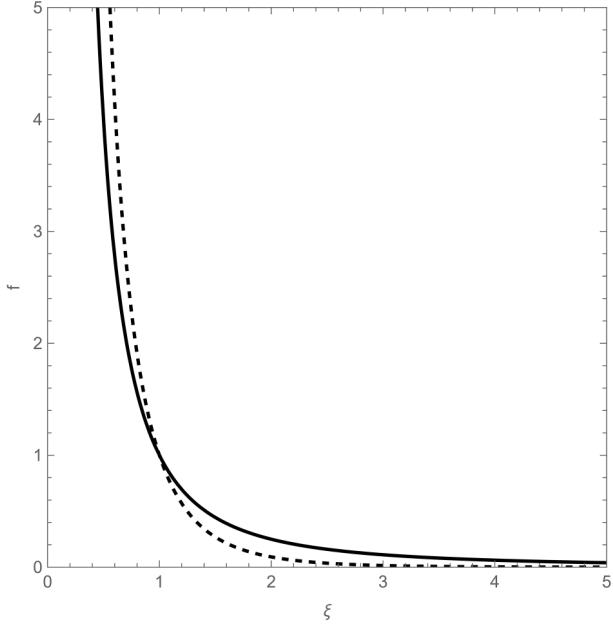

**Figure 12.** Functional form of $f_G$ (—) and $f_R$ (- - - -) for a static equilibrium configuration at $\bar{\xi} = 1$.

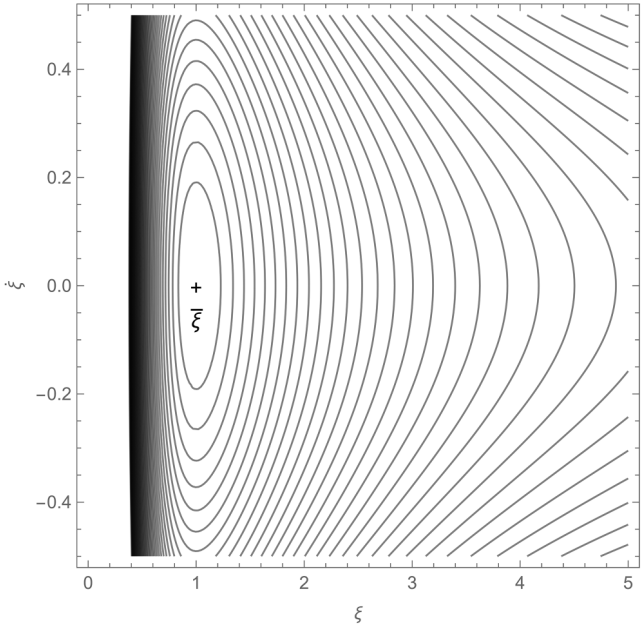

**Figure 13.** Phase space with stable equilibrium configuration for Dyson bubble element at $\bar{\xi} = 1$.

It can be seen from equation (46) that the transverse motion of the reflector is also passively stable. It can therefore be concluded that, to linear order, a Dyson bubble can in principle be self-stabilizing when the cloud of reflectors is sufficiently dense to generate mutual attenuation and so alter the force law for radiation pressure. This implies that Dyson bubbles could in principle be long-lived structures. The implications for technosignatures will be discussed further in Section 7.2.

### 6.5 Self-gravity of the Dyson bubble and scattered radiation

In order to assess the assumptions of the simple model used in Section 6.4, the self-gravity of the Dyson bubble will now be considered. If the Dyson bubble is assumed to a homogeneous dense cloud then Newton's shell theorem can be utilized (B. C. Reed 2022). First, consider a reflector located within the Dyson bubble at some radial distance $r$ from the central star. From Newton's shell theorem the reflector will experience a gravitational force from the remainder of the reflectors located in the sphere interior to its location, as if that mass were concentrated as a single point. There is no net gravitational force due to the reflectors located in the volume exterior to its location, again according to Newton's shell theorem. For a Dyson bubble of number density $n$ and reflector mass $m$, the mass of the Dyson bubble interior to the reflector located at distance $r$ is given by $(4\pi/3)\,nm\,r^3$. The total gravitational acceleration experienced by a single reflector, from the central star and the Dyson bubble, is then $GM_*/r^2 + (4\pi G/3)nmr$, directed radially inwards towards the central star.

The effect of light scattered from reflectors within the Dyson bubble can also be considered. If the Dyson bubble is again considered as a homogeneous dense cloud, a diffuse and uniform background of scattered radiation will be assumed. Indeed, for an optically thick medium, the specific intensity of a radiation field can be approximated solely by its source function (G. B. Rybicki and A. P. Lightman 2004). Clearly a uniform diffuse background of scattered radiation is a simplification, however it will illustrate how the Dyson bubble can be destabilized under certain conditions. For a reflector with dissimilar optical properties, such as a reflective front surface and absorbing rear surface, in very general terms a constant radiation pressure force can be considered to model the effect of the diffuse background of scattered radiation. This diffuse background is therefore assumed to generate an additional constant acceleration $\varphi$. The total radiation pressure acceleration experienced by a single reflector, from the central star and the Dyson bubble, is then $\beta GM_*e^{-nAr}/r^2 + \varphi$, directed radially outwards away from the central star.

In non-dimensional units, incorporating the mass of the Dyson bubble and the simple model of the diffuse background of scattered radiation, the dynamics of a reflector can now be defined by

$$\ddot{\xi} = -\Omega^2 \frac{1}{\xi^2} - \Omega^2 \lambda_m \xi + \beta \Omega^2 \frac{1}{\xi^2} e^{-\mu\xi} + \Omega^2 \lambda_r \tag{47}$$

where $\lambda_m = 4\pi nmR^3/3M_*$ and $\lambda_r = \varphi R^2/GM_*$. For a Dyson bubble the parameter $R$ is now an arbitrary normalization length-scale, since the reflectors are considered to be small. Therefore, if $R$ is now considered to be the length-scale of the Dyson bubble, with $\Omega^2 = GM_*/R^3$, it can be seen that $\lambda_m$ scales the ratio of the mass of the Dyson bubble to the mass of the central star. Similarly, it can be seen that $\lambda_r$ scales the ratio of the radiation pressure acceleration from the diffuse background radiation to the gravitational acceleration from the central star at the edge of the Dyson bubble, which is equivalent to the radiation pressure acceleration from the central star if $\bar{\beta}_D \sim 1$.

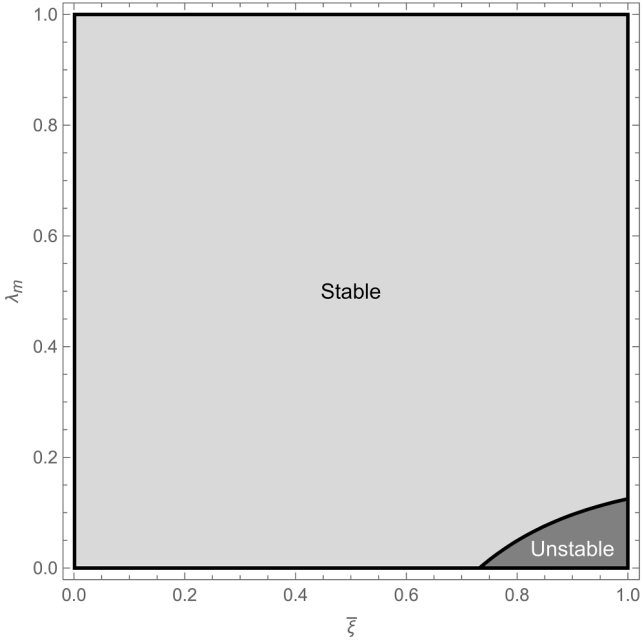

**Figure 14.** $\bar{\xi} - \lambda_m$ parameter space with stable (light shaded) and unstable (dark shaded) regions.

For static equilibrium at $\xi = \bar{\xi}$ the required lightness number $\bar{\beta}_D$ can be determined from equation (47) and is found to be $\bar{\beta}_D = e^{\mu \bar{\xi}}(1 + \lambda_m \bar{\xi}^3 - \lambda_r \bar{\xi}^2)$. Then, following the linearization procedure of equation (43) with the modified gravitational and radiation pressure acceleration from equation (47), it can be shown that

$$\left(\frac{\Lambda}{\Omega}\right)^2 = \mu \frac{1}{\bar{\xi}^2} + \lambda_m \left(3 + \mu \bar{\xi}\right) - \lambda_r \left(\mu + \frac{2}{\bar{\xi}}\right). \quad (48)$$

First, it can be seen that if $\lambda_r = 0$ then $(\Lambda/\Omega)^2 > 0$ for all $\lambda_m$ and so the Dyson bubble is always stable if its self-gravity is considered and the scattered radiation neglected. It can also be noted that $\lambda_r = 0$ if the scattered radiation does not generate a net acceleration, for example for a cloud of optically symmetric absorbing elements. Indeed, it can also be seen that if $\lambda_r = 0$ and $\mu = 0$ then $(\Lambda/\Omega)^2 = 3\lambda_m$ and so $(\Lambda/\Omega)^2 > 0$. This highlights an important alternative passive stabilization mechanism for a Dyson bubble: if the Dyson bubble has mass, then the deviation of the gravitational force from an inverse square law due to the self-gravity of the Dyson bubble is in itself sufficient to ensure passive stability. Since the transverse dynamics of the problem are uncoupled to linear order, as demonstrated in Appendix C, the transverse motion of the reflector will remain stable.

Moreover, if $\lambda_m = 0$ it can be seen that $(\Lambda/\Omega)^2 > 0$ if $\lambda_r < \mu/\bar{\xi}(2 + \mu\bar{\xi})$, and so the Dyson bubble will remain stable if only the diffuse background radiation is considered and the resulting radiation pressure is bounded from above. An example of this condition is shown in Fig. 14 for $\lambda_r = 0.5$ and $\mu = 1$ for illustration, where the inequality maps onto the unstable region of the parameter space, partitioned by $\bar{\xi} = 0.527$ when $\lambda_m = 0$. It can also be seen in Fig. 14 that for $\lambda_m \ll 1$ the diffuse background radiation destabilizes the Dyson bubble at its outer edge, as the radiation pressure from the central star diminishes. However, for $\lambda_m \not\to 0$ the mass fraction of the Dyson bubble has a stabilizing effect, as expected from equation (48).

For $\lambda_m \not\to 0$ and $\lambda_r \not\to 0$ it can be shown that $(\Lambda/\Omega)^2 > 0$ if $\lambda_r < (\mu + \lambda_m \bar{\xi}^2(3 + \mu \bar{\xi}))/\bar{\xi}(2 + \mu \bar{\xi})$, and again for stability the radiation pressure from the diffuse background must be bounded from above. However, for a dense cloud such that $\mu \gg 1$, it can be shown that $\lambda_r \lesssim (1 + \lambda_m \bar{\xi}^3)/\bar{\xi}^2$. At the edge of the Dyson bubble $\bar{\xi} \cong 1$ and so $\lambda_r \lesssim 1 + \lambda_m$. Since $\lambda_m > 0$, and it can be expected that $\lambda_r \lesssim 1$, then $(\Lambda/\Omega)^2 > 0$ and so the Dyson bubble will be stable.

Using very general considerations, it has been shown that the Dyson bubble can remain stable when its self-gravity and a simple model of a diffuse background of scattered radiation are included in the dynamics defined in Section 6.4. However, there are now regions of the parameter space where instability can occur, primarily at the edge of the Dyson bubble driven by the diffuse background radiation. In addition, it has been shown that the self-gravity of the Dyson bubble is in itself sufficient to ensure passive stability in the absence of the diffuse background radiation, and indeed it enhances the stability of the Dyson bubble when the diffuse background of scattered radiation is included.

## 7. IMPLICATIONS FOR TECHNOSIGNATURES

Now that the stability properties of the stellar engine, Dyson bubble and Dyson swarm have been investigated, the implications of these properties for technosignatures can be assessed. Arguably, it can be imagined that such structures would be configured to be in passively stable configurations. Therefore, understanding the stable region of the parameter space can in principle assist SETI searches by constraining the geometries and configurations of candidate objects. Moreover, passively stable structures offer the possibility of relic objects surviving from the past with no intervention.

### 7.1 Stellar engines

It has been demonstrated in Section 6.2 that while an ideal stellar engine comprising a uniform reflective rigid disc is unstable, a reflective disc whose mass is concentrated at the edge of the disc can in principle be passively stable. Moreover, the analysis demonstrated that the limiting case for a stable configuration occurs when the disc is in equilibrium at a non-dimensional distance of $\bar{\xi} = 1/\sqrt{2}$ from the central star. Moreover, the maximum gravitational force exerted on the ring occurs when $df_G/d\xi = 0$, which occurs at the stability boundary when $\bar{\xi} = 1/\sqrt{2}$, as can be shown from equation (5) (and Fig. 10). This will generate the maximum propulsive force for a stellar engine, since the radiation pressure force must balance the gravitational force. From equation (10) the resulting maximum radiation pressure force exerted on the reflector is then $L_*/3c$, which provides an upper bound on the acceleration of the star which can be achieved of $L_*/3M_*c$. Similarly, the portion of the stellar energy output energy intercepted by the reflector can be determined from the fractional solid angle subtended by the disc and is found to be $1/2(1 + \bar{\xi}^2)$. This is simply $1/3$ when $\bar{\xi} = 1/\sqrt{2}$, as expected from the maximum radiation pressure force. The reflector, therefore, subtends a half-angle of approximately 55° at the central star in thse limiting stable configuration.

Since the specific configuration where $\bar{\xi} = 1/\sqrt{2}$ corresponds to the limit of stability, a stable stellar engine which can maximise the acceleration of the star is expected at $\bar{\xi} \gtrsim 1/\sqrt{2}$. However, for a fixed stellar luminosity, a slow degradation of the reflectivity of the stellar engine would lead to a continuous decrease of the equilibrium displacement $\bar{\xi}$, potentially past the stability boundary. The impact of increasing stellar luminosity, which can counteract such degradation, will be discussed below. In summary, a passively stable stellar engine appears able to command up to 1/3 of the energy and momentum flux from the central star, arguably resulting in a distinctive technosignature (D. H. Forgan 2013; M. Lingam & A. Loeb 2020).

### 7.2 Dyson bubbles

A self-stabilizing static Dyson bubble would appear as a dense cloud enclosing the central star. This would modify the spectral characteristics of the central star, with an infrared excess as also expected for a solid Dyson sphere. Moreover, for a static cloud there would be no flickering of the star for a distant observer, unlike a swarm of orbiting reflectors which would pass in front of the stellar disc. A self-stabilizing static cloud would also avoid or minimize collisions between the elements of the cloud, which would be an issue for an orbiting Dyson swarm with a large number of small discs (B. C. Lacki 2025). There would also be no velocity-dependent Poynting–Robertson forces to perturb the long-term dynamics of the reflectors. However, the cloud would still be subject to gravitational perturbations due to distant stellar encounters, while interstellar comets passing close to the central star would pass through the cloud itself and would pose a collision risk.

Lacki also noted that the evolution of the central star's luminosity will perturb radiation pressure supported reflectors, essentially destroying the equilibria (B. C. Lacki 2025). Due to the slowly increasing luminosity of a Sun-like central star as it ages, the effective lightness number of the reflector will increase since it scales as $L_*/M_*$. Consider first the case of inverse square radiation pressure defined by equation (39), with a lightness number now defined by $\beta = 1 + \delta\beta(t)$ where $\delta\beta(t) > 0$. Here, $\beta = 1$ is the lightness number required for static equilibrium and $\delta\beta(t)$ scales as the slowly increasing stellar luminosity. It can be seen that the nett acceleration experienced by a reflector is now $(\Omega^2/\xi^2)\delta\beta(t) > 0$. Since in this case the inverse square radiation pressure force is exactly balanced the inverse square gravitational force at all distances from the central star, any increase in stellar luminosity, and hence lightness number, will disrupt the Dyson bubble.

However, for a dense Dyson bubble a slow increase in stellar luminosity will displace rather than destroy equilibria. This can be seen from Fig. 12, where an increase in the force due to radiation pressure will displace the equilibrium point to a greater distance from the central star. From equation (42) it was shown that the required lightness number for a Dyson bubble is given by $\bar{\beta}_D = e^{\mu\bar{\xi}}$. Therefore, if the stellar luminosity, and hence the lightness number, slowly increases the equilibrium position of the reflector will change quasi-statically such that $\bar{\xi}(t) \sim \log\bar{\beta}_D(t)/\mu(t)$, where the optical depth will also slowly decrease due to the expansion of the bubble. Since $\bar{\beta}_D(t)$ and $\mu(t)$ are always positive, in principle the equilibrium distance $\bar{\xi}(t)$ remains finite, indicating robustness for the dense Dyson bubble.

The same robustness can also be seen in the ring stellar engine. From equation (34) it was shown that the required lightness number for static equilibrium is given by $\bar{\beta}_R = \bar{\xi}/\sqrt{1+\bar{\xi}^2}$. Again, if the stellar luminosity, and hence the lightness number, slowly increase the equilibrium position of the reflector will change quasi-statically such that $\bar{\xi}(t) \sim \bar{\beta}_R/\sqrt{1-\bar{\beta}_R^2}$. For an equilibrium configuration close to the stable boundary where $\bar{\xi}(t) \sim 1/\sqrt{2}$ the required lightness number $\bar{\beta}_R \sim 1/\sqrt{3}$. However, it is not until the lightness number $\bar{\beta}_R \to 1$ that $\bar{\xi} \to \infty$ and the equilibrium configuration vanishes. Therefore, the ring stellar engine appears to be robust to a stellar luminosity increase of $\sqrt{3}-1$.

Speculatively, the optical properties of the reflectors for the Dyson bubble could be engineered to degrade to match the slowly increasing luminosity of a Sun-like central star as it ages. Moreover, for perfectly absorbing discs for a Dyson bubble, which can be envisaged for energy collection and processing, the optical properties of the reflectors would be fixed, since an initially perfectly absorbing disc cannot further degrade in reflectivity. Moreover, if the optical properties of discs are also symmetric then re-radiated energy would not generate a perturbation due to radiation pressure.

In summary, the self-stabilizing nature of the dynamics of dense Dyson bubbles, and their apparent robustness, in principle implies an inherent longevity for such ultra-large space structures. Therefore, in addition to a Dyson bubble which is in active use, observation of such structures could also indicate relic objects which survive from the distant past to the present without active intervention.

### 7.3 Dyson swarms

As an alternative to a static equilibrium configuration, a Dyson swarm can be envisaged with a number of reflectors in orbit about a central star, again either a small number of ultra-large reflective discs with $R \gg R_*$, or a large number of small reflective discs with $R_* \gg R$. Then, the stability criteria defined by equations (28) and (31) can be used to assess orbit stability when $R \gg R_*$ and $R_* \gg R$, repectively. The analysis suggests that a Dyson swarm should be assembled from either a small number of large discs orbiting in the stable region of the $\xi-\beta$ parameter space defined by $\beta < \bar{\beta}_S$ (Fig. 7) or a large number small discs orbiting in the stable region of the $\zeta-\beta$ parameter space defined by $\beta < \bar{\beta}_S^*$ (Fig. 9). A Dyson swarm of orbiting discs can therefore ensure passive stability when configured to be within the correct region of the appropriate parameter space. Again, passive stabilization with the reflector normal always pointing away from the central star can be envisaged using slighting conical reflectors with the centre-of-pressure displaced behind the centre-of-mass.

A Dyson swarm can be expected to generate a different technosignature to a passively stable Dyson bubble discussed above. For example, the motion of the discs in a swarm would imply a flickering of the observed luminosity of the central star, with a larger variation expected from a small number of ultra-large discs relative to a large number of small discs. Finally, while an orbiting swarm of reflectors will be susceptible to collisions (B. C. Laki 2025), collisions within a Dyson swarm could in principle be minimised using families of displaced non-Keplerian orbits, where the orbit planes of the reflectors can be stacked in parallel rather than being inclined relative to each other (C. R. McInnes & J. F. L. Simmons 1992).

### 7.4 Implications of zero-stress discs for SETI

Aspects of the analysis presented in this paper have assumed perfectly reflecting, ultra-large rigid discs where $R \gg R_*$ in order to obtain new insights into the orbital dynamics of stellar engines and Dyson swarms. However, realistic reflectors will have internal stresses due to the gradient of gravitational and radiation pressure forces across the disk. Appendix A has demonstrated that an ultra-large zero-stress disc can be configured if the disc is rotating as a solid body about its principal axis and its local lightness number and reflectivity vary across the disc from its centre to its edge. Moreover, a non-rotating disc can be in a zero-stress configuration if it is perfectly absorbing and its local lightness number varies across the disc. A perfectly reflecting zero stress disc is only possible if the elements move as independent bodies. It can also be noted that the force laws derived scale using a non-dimensional independent variable $\xi = r/R$. Therefore, as $R \to 0$ then $\xi \to \infty$ for a fixed physical distance $r$ and so the force laws are in principle applicable to a range of disc sizes, although the assumption in Sections 2 and 3 is that the central star is point-like relative to the disc.

In terms of applications, a rotating zero-stress disc can be envisaged as a configuration suitable for a static stellar engine, while a non-rotating disc can be considered for applications requiring circular orbits such as the Dyson swarm. An orbiting rotating disc will experience coupling between the rotation of the disc and orbital motion of the disc about the central star, which is not considered here. Similarly, an orbiting non-rotating disc will experience stress due to the differential centripetal force acting across the disc due to its motion about the central star. Again, this is not considered here but could be added to the analysis of Appendix A. Perfectly absorbing discs can be envisaged for Dyson swarms where energy is to be collected from the central star.

It can be noted that the Dyson bubble configuration assumes small discs where $R_* \gg R$, as can the Dyson swarm, and so these configurations are not subject to the same issues associated with ultra-large discs. Ultra-large discs which have a local lightness number and reflectivity which varies across the disc will have a different force law to the ideal discs considered here, which can be investigated in future work. However, the simplified models presented in this paper provide insights into how passive stability can be engineered, to begin to understand the requirements for long-lived ultra-large space structures.

## 8. CONCLUSIONS

It has been demonstrated using a simple model that disc stellar engines and Dyson bubbles are unstable due to the functional form of the gravitational and radiation pressure forces exerted on such structures. However, strategies have been presented which can enable passively stable stellar engines and self-stabilizing Dyson bubbles. A stellar engine can in principle be stabilized using a ring configuration while a Dyson bubble can in principle be stabilized if a vast number of reflectors are deployed in a dense cloud. The self-gravity of the cloud can also ensure stability. Moreover, the stability properties of orbiting Dyson swarms have been investigated and stable regions of the parameter space identified. The condition for equilibrium for a non-homogenous, partially reflecting rotating disc which can enable an ultra-large zero-stress reflector has also been investigated. Understanding the conditions required for passive stability of such ultra-large space structures, and how to engineer such configurations, can provide insights into the likely characteristics of technosignatures in search for SETI studies. Speculatively, passive stability may also enable relic structures to survive from the distant past without active intervention.

**ACKNOWLEDGEMENTS**

This work was supported by the Department of Science, Innovation and Technology (DSIT) and the Royal Academy of Engineering under the Chair in Emerging Technologies programme. For the purpose of open access, the author has applied a Creative Commons Attribution (CC-BY) licence to any Author Accepted Manuscript version arising from this submission.

**DATA AVAILABILITY**

The data that support the findings of this study are available from the corresponding author upon reasonable request. Computations were performed using Mathematica 14.2.

**REFERENCES**

Badescu V., Cathcart R. B., 2006, Acta Astron, 58, 119
Borgraffe A., Heiligers J., Ceriotti M., McInnes C. R., 2014, IAC-14.C1.3.4, 65th IAC Cong. Toronto, Canada
Borgraffe A., Heiligers J., Ceriotti M., McInnes C. R., 2015, Proc. R. Soc. A, 471, 2015119
Caplan M. E., 2019, Acta Astron, 165, 96
Drexler K. E., 1979, MSc Thesis, Massachusetts Institute of Technology
Dyson F. J., 1960, Science, 131, 1667
Forgan D. H., 2013, J. Br. Interplanet. Soc., 66, 144
Harwit M., 2006, Astrophysical Concepts, Springer, New York
King A. C., Billingham J., Otto R. S., 2003, Differential Equations, Linear, Nonlinear, Ordinary, Partial, Cambridge University Press, Cambridge
Lacki B. C., 2016, preprint (arXiv:1604.07844)
Lacki B. C., 2025, ApJ, 985, 191
Lingham M., Loeb A.,2020, ApJ, 905, 175
Mahomed F. M., Vawda F., 2000, Nonlinear Dyn., 21, 307
Maxwell J. C., 1859, On the Stability of the Motion of Saturn's Rings, McMillian, Cambridge
McInnes C. R., 1991, PhD Thesis, Univ. Glasgow
McInnes C. R., 1999, J. Guid. Control Dyn., 22, 185
McInnes C. R., 2002, Ap&SS, 282, 765
McInnes C. R., 2004, Solar Sailing: Technology, Dynamics and Mission Applications, Springer, Berlin
McInnes C. R., 2007, J. Guid. Control Dyn, 30, 870
McInnes C. R., 2009, in Badescu V., ed, Mars: Prospective Energy and Material Resources, Springer, Berlin
McInnes C. R., 2025, MNRAS, 537, 1249
McInnes C. R., Brown J. C., 1990, Celest. Mech. Dyn. Astro., 49, 249
McInnes C. R., Simmons J. F. L., 1992, J. Spacecr. Rockets, 29, 466
Reed B. C., 2022, Am. J. Phys., 90, 394
Rybicki G. B., Lightman A. P., 2004, Radiative Processes in Astrophysics, Wiley-VCH, Weinheim
Sánchez Cuartielles J. P., McInnes C. R., 2015, PLoS One, 10, e0136648
Schumayer D., Hutchinson D. A. W., 2019, Am. J. Phys., 87, 384
Seifritz W., 1989, Nature, 340, 603
Shkadov L. M., 1987, IAA-87–613, 38th IAF Cong., Brighton, UK
Suazo M. et al., 2022, MNRAS, 512, 2988
Suazo M. et al., 2024, MNRAS, 531, 695
Svoronos A. A., 2020, Acta Astron, 176, 306
Swartzlander G. A., 2022, J. Opt. Soc. Am. B, 39, 2556
Wright J. T., 2020, Serb. Astron. J., 200, 1
Wright J. T. et al., 2022, ApJ, 927, L30
Zubrin R., McKay C. P., 1993, AIAA-93-2005, AIAA, SAE, ASME, and ASEE, 29th JPC, Monterey, USA

## APPENDIX A: EQUILIBRIUM STATE FOR A NON-HOMOGENOUS, PARTIALLY REFLECTING ROTATING DISC

In order to assess if a zero-stress reflector can in principle be configured, the gravitational and radiation pressure forces acting on an infinitesimal element of the reflector will be considered. In particular, a partially reflecting disc will be considered since a component of the radiation pressure force will now be generated within the plane of the reflector. Moreover, the reflector will now be allowed to rotate about its principal axis of symmetry so that there is a spin-induced in-plane force. In related work, it has been shown that a parabolic reflector can be formed through elastic deformation of a thin rotating disc using a reflectivity profile which varies radially across the reflector (A. Borgraffe et al. 2015). Recent developments in metamaterials have also been considered to engineer radiation pressure forces in reflectors (G. A. Swartzlander 2022), while distributed reflective elements with locally time-varying properties have been considered for active attitude control (A. Borgraffe et al. 2014). For ultra-large structures it has also been speculated that active control of monolithic Dyson spheres could delay the onset of bucking modes through the use of embedded sensors and smart materials (C. R. McInnes 2025). However, rather than active control, a passive strategy will now be considered to generate a zero-stress reflector.

To ensure that each infinitesimal element of the reflector is in equilibrium, its reflectivity and the areal density (and hence local lightness number) will be allowed to vary radially from the centre of the disc. The gravitational force $d\boldsymbol{f}_G$ acting on a mass element $dm$ can be defined using equation (1) in terms of a normal component in direction $\boldsymbol{n}$, defined in Fig. 1, and a transverse component $\boldsymbol{t}$, where the unit vector $\boldsymbol{t}$ is in direction $\boldsymbol{\rho}$ along the plane of the reflector. Moreover, the centripetal induced in-plane force will be appended to the gravitational force, where the disc rotates as a solid body with a fixed angular velocity $\Omega$ so that:

$$d\boldsymbol{f}_G = -\frac{GM_*}{d^2}(\boldsymbol{u}.\boldsymbol{n})dm\,\boldsymbol{n} - \frac{GM_*}{d^2}(\boldsymbol{u}.\boldsymbol{t})dm\,\boldsymbol{t} + \Omega^2\rho\,dm\,\boldsymbol{t} \quad (A1a)$$

Then, the radiation pressure force $d\boldsymbol{f}_R$ acting on an infinitesimal area element $dA$ can be written as (C. R. McInnes 1999)

$$d\boldsymbol{f}_R = \frac{1}{2}\beta\frac{GM_*}{d^2}(1+\varepsilon)(\boldsymbol{u}.\boldsymbol{n})^2\sigma dA\,\boldsymbol{n}$$
$$+ \frac{1}{2}\beta\frac{GM_*}{d^2}(1-\varepsilon)(\boldsymbol{u}.\boldsymbol{n})(\boldsymbol{u}.\boldsymbol{t})\sigma dA\,\boldsymbol{t} \quad (A1b)$$

where $\varepsilon$ is the local reflectivity and $\beta$ is the local lightness number. It can be seen that for a partially reflectivity with $\varepsilon < 1$ there will now be a transverse component of the radiation pressure force. If it is assumed that the absorbed energy is uniformly radiated from both sides of the reflector then the only nett radiation pressure force is due to incident and reflected photons. Defining a defining non-dimensional distances $\xi = r/R$ and $\varsigma = \rho/R$, where $0 \leq \varsigma \leq 1$, it can be seen that

$$d\boldsymbol{f}_G = -\bar{a}\frac{\xi}{(\xi^2+\varsigma^2)^{3/2}}\sigma dA\,\boldsymbol{n} - \bar{a}\frac{\varsigma}{(\xi^2+\varsigma^2)^{3/2}}\sigma dA\,\boldsymbol{t}$$
$$+ \bar{\Omega}^2\varsigma\,\sigma dA\,\boldsymbol{t} \quad (A2a)$$

$$d\boldsymbol{f}_R = \frac{1}{2}\beta\bar{a}\frac{\xi^2}{(\xi^2+\varsigma^2)^2}(1+\varepsilon)\sigma dA\,\boldsymbol{n} + \frac{1}{2}\beta\bar{a}\frac{\xi\varsigma}{(\xi^2+\varsigma^2)^2}$$
$$\times (1-\varepsilon)\sigma dA\,\boldsymbol{t} \quad (A2b)$$

where $\bar{a} = GM_*/R^2$, $\bar{\Omega} = \sqrt{R}\Omega$, and $dm = \sigma dA$. For each infinitesimal mass element to be in equilibrium clearly the condition $d\boldsymbol{f}_R + d\boldsymbol{f}_G = 0$ is required at all points across the reflector. Then, the force components in the normal and transverse directions can be equated in order to generate an equilibrium state for each infinitesimal mass element. If each mass element is in equilibrium independently then there will be no differential forces and hence no mechanical stress within the reflector. It will now be assumed that the reflector is at some radial distance $\xi = \bar{\xi}$ from the central star. Then, a rotating reflector with $\bar{\Omega} \neq 0$ will be considered with a reflectivity and lightness number which varies radially across the reflector. By equating the normal force components from equations (A2) it is found that

$$\beta(\varsigma) \triangleq \frac{\sigma^*}{\sigma(\varsigma)} = \frac{2}{1+\varepsilon(\varsigma)}\sqrt{1+\left(\frac{\varsigma}{\bar{\xi}}\right)^2} \quad (A3)$$

so that the lightness number $\beta(\varsigma)$, and hence the local areal density $\sigma(\varsigma)$, is a function of the radial distance along the reflector from its centre to its edge, again where $0 \leq \varsigma \leq 1$. Then, using equation (A3) and equating the transverse force components in equations (A2), it can be shown that

$$\varepsilon(\varsigma) = \frac{(\bar{\Omega}/\Omega_*)^2}{2-(\bar{\Omega}/\Omega_*)^2} \quad (A4)$$

where $\Omega_*^2 = \bar{a}/(\bar{\xi}^2+\varsigma^2)^{3/2}$ and $0 \leq \varepsilon(\varsigma) \leq 1$. Therefore, for a fixed reflector angular velocity $\bar{\Omega}$, the local reflectivity must also be a function of the radial distance along the reflector. If the boundary condition $\varepsilon(1) = 1$ is set, so that there is perfect reflectivity at its edge, then from equation (A4) it can be seen that $\bar{\Omega}/\Omega_* = 1$ and so the required reflector angular velocity is given by $\bar{\Omega}^2 = \bar{a}/(1+\bar{\xi}^2)^{3/2}$. For illustration it will be assumed that $\bar{a} = 1$ and $\bar{\xi} = 1$ so that $\bar{\Omega}^2 = 2^{-3/2}$. The reflectivity at the centre of the disc is then $\varepsilon(0) = 0.215$ while the lightness number has boundary conditions $\beta(0) = 1.646$ and $\beta(1) = 1.414$, as shown in Fig. A1.

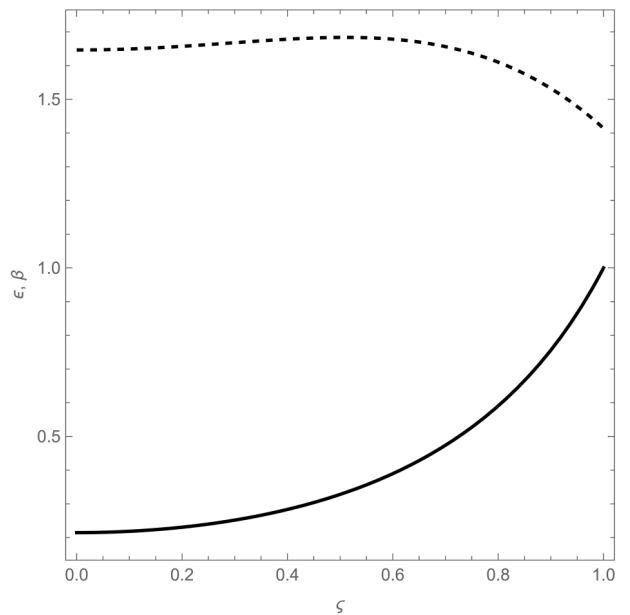

**Figure A1.** Functional form of $\varepsilon(\varsigma)$ (—) and $\beta(\varsigma)$ (- - - -) with $\varepsilon(1) = 1$.

For a non-rotating reflector with $\bar{\Omega} = 0$ it can be seen from equation (A4) that equilibrium can only be achieved if the reflectivity $\varepsilon = 0$ and the lightness number varies across the reflector such that $\beta(\varsigma) = 2\sqrt{1 + (\varsigma/\bar{\xi})^2}$. Therefore, a zero-stress reflector is in principle possible, however it must be optically absorbing. Again, if it is assumed that the absorbed energy is uniformly radiated from both sides of the reflector then the only nett radiation pressure force is due to incident photons. Moreover, for a uniform ideal reflector with $\varepsilon(\varsigma) = 1$ for $0 \leq \varsigma \leq 1$ it can be seen from equation (A4) that the condition $\bar{\Omega} = \Omega_*$ is required. However, for a continuous reflector $\bar{\Omega}$ is fixed while $\Omega_*$ is a function of $\varsigma$ for a fixed $\bar{\xi}$. Therefore, in this specific case equilibrium is only possible if the infinitesimal elements move as independent bodies. This corresponds to the reflectors moving on displaced circular orbits, levitated above the central star and are equivalent to the family of so-called Keplerian synchronous mode non-Keplerian orbits for solar sails (C. R. McInnes and J. F. L. Simmons 1992).

## APPENDIX B: STABILITY OF A DISPLACED STELLAR RING

In order to determine the full stability properties of a stellar ring, an infinitesimal transverse displacement $\delta\eta$ will now be applied to the ring, orthogonal to the $\xi$-axis. From Fig. 1, the position of the centre of the ring can now be written in dimensional coordinates as $\boldsymbol{r} = (r, 0, \delta z)$ where $\delta z = R\delta\eta$ for a ring of radius $R$. The unit vector $\boldsymbol{u}$ can then be defined as $\boldsymbol{d}/\|\boldsymbol{d}\|$ where $\boldsymbol{d} = \boldsymbol{r} + \boldsymbol{\rho}$ and $\boldsymbol{\rho} = (0, R\cos\phi, R\sin\phi)$. It will be assumed that the reflector normal is fixed in an inertial frame of reference such that $\boldsymbol{n} = (1, 0, 0)$, for example through passive spin stabilization, so that the radiation pressure force will remain directed along the $\xi$-axis. Moreover, since the transverse perturbation is infinitesimal, it will be assumed that there are no gradient torques acting on the ring or disc. The radial coordinate will be left as arbitrary so that a simultaneous radial and transverse displacement can be considered. It is found that to linear order the effect of both displacements are uncoupled, so that linearizing both the radial and transverse dynamics leads to equation (35) to describe the linearised radial dynamics of the problem.

Therefore, from equation (1) and using the new definitions of $\boldsymbol{d}$ and $\boldsymbol{u}$ it can be shown that gravitational force $d\boldsymbol{f}_G$ exerted on the mass element $dm$ can be now written as

$$d\boldsymbol{f}_G = -GM_* \begin{pmatrix} r \\ R\cos\phi \\ \delta z + R\sin\phi \end{pmatrix} \times \frac{dm}{\left(r^2 + R^2\cos^2\phi + (\delta z + R\sin\phi)^2\right)^{3/2}} \quad \text{(B1)}$$

Then, re-arranging an expanding equation (B1) in powers of $(\delta z/r)$ it can be shown that

$$d\boldsymbol{f}_G = -\frac{GM_*}{(R^2 + r^2)^{3/2}} \begin{pmatrix} r \\ R\cos\phi \\ \delta z + R\sin\phi \end{pmatrix} \times \left(1 - 3\frac{(R/r)(\delta z/r)}{1 + (R/r)^2}\sin\phi + O\left((\delta z/r)^2\right)\right) \lambda d\phi \quad \text{(B2)}$$

where $\lambda$ is the linear density of the ring. The total gravitational force $\boldsymbol{f}_G$ can then be determined by integration around the ring

such that

$$\boldsymbol{f}_G = -\frac{GM_*\lambda}{(R^2 + r^2)^{3/2}} \int_0^{2\pi} \begin{pmatrix} r \\ R\cos\phi \\ \delta z + R\sin\phi \end{pmatrix} \times \left(1 - 6\frac{(R/r)(\delta z/r)}{1 + (R/r)^2}\sin\phi + O\left((\delta z/r)^2\right)\right) d\phi \quad \text{(B3)}$$

Upon integration over $\phi$ the trigonometric terms vanish and the integral then reduces to

$$\boldsymbol{f}_G = -\frac{GM_*m}{R^2} \begin{pmatrix} \frac{(r/R)}{\left(1+(r/R)^2\right)^{3/2}} \\ 0 \\ \frac{1}{2}\frac{2(r/R)^2-1}{\left(1+(r/R)^2\right)^{5/2}}\frac{\delta z}{R} \end{pmatrix} \quad \text{(B4)}$$

It can be seen that the radial component of the gravitational force exerted on the ring remains unchanged, while to linear order there is a transverse restoring force proportional to $\delta z$. This is to be expected. If the position of the ring is displaced by some small angle $\delta\gamma$ away from the radial direction as seen from the central star, then the gravitational force will have a radial component $-f_G\cos\delta\gamma \approx -f_G$ and a transverse component $-f_G\sin\delta\gamma \approx -f_G\delta\gamma$. It can be seen that it can be expected that the radial force remains unchanged while there is a stable transverse restoring force, as demonstrated by equation (B4).

Now that the gravitational force on the ring has been determined the radiation pressure force exerted on the reflective disc can be calculated. As above, the radial coordinate will be left as arbitrary so that a simultaneous radial and transverse displacement can be considered. Linearizing both the radial and transverse dynamics leads to equation (35) to describe the linearized radial dynamics of the problem. Using the definition $\boldsymbol{u} = \boldsymbol{d}/\|\boldsymbol{d}\|$ where $\boldsymbol{d} = \boldsymbol{r} + \boldsymbol{\rho}$, and again fixing $\boldsymbol{n} = (1, 0, 0)$, from equation (7) the radiation pressure force $d\boldsymbol{f}_R$ exerted on the area element $dA$ can now be written as

$$d\boldsymbol{f}_R = \frac{2}{c}\frac{L_*}{4\pi}\begin{pmatrix}1\\0\\0\end{pmatrix}\frac{r^2\rho d\rho d\phi}{\left(r^2 + \rho^2\cos^2\phi + (\delta z + \rho\sin\phi)^2\right)^2} \quad \text{(B5)}$$

Then, re-arranging an expanding equation (B5) in powers of $(\delta z/r)$ it can be shown that

$$d\boldsymbol{f}_R = \frac{2}{c}\frac{L_*}{4\pi}\frac{r^2}{(\rho^2 + r^2)^2}\begin{pmatrix}1\\0\\0\end{pmatrix} \times \left(1 - 4\frac{(\rho/r)(\delta z/r)}{1 + (\rho/r)^2}\sin\phi + O\left((\delta z/r)^2\right)\right)\rho d\rho d\phi \quad \text{(B6)}$$

The total radiation pressure force $\boldsymbol{f}_R$ can then be determined by integration across the disc such that

$$\boldsymbol{f}_R = \frac{2}{c}\frac{L_*}{4\pi}r^2 \int_0^{2\pi}\int_0^R \frac{1}{(\rho^2 + r^2)^2}\begin{pmatrix}1\\0\\0\end{pmatrix} \times \left(1 - 4\frac{(\rho/r)(\delta z/r)}{1 + (\rho/r)^2}\sin\phi + O\left((\delta z/r)^2\right)\right)d\rho d\phi \quad \text{(B7)}$$

Again, upon integration over $\phi$ the trigonometric term vanishes and the force defined by equation (9) is recovered such that

$$\boldsymbol{f}_R = \frac{L_*}{2c}\begin{pmatrix} 1 - \frac{(r/R)^2}{1+(r/R)^2} \\ 0 \\ 0 \end{pmatrix} \quad \text{(B8)}$$

To linear order the magnitude of the radiation pressure force exerted on the ring therefore remains unchanged due to the displacement $\delta z$. Again, if the position of the reflector is displaced by some small angle $\delta \gamma$ away from the radial direction as seen from the central star, then the radiation pressure force will scale as $f_R \cos^2 \delta \gamma \approx f_R$ if $\delta \gamma \ll 1$. Therefore, it is expected that to first order the radiation pressure force exerted on the ring remains unchanged, as demonstrated by equation (B8).

## APPENDIX C: STABILITY OF A DISPLACED DYSON BUBBLE REFLECTOR

In order to determine the full stability properties of a single reflector in a Dyson bubble, an infinitesimal transverse displacement $\delta \eta$ will now be applied to the reflector, orthogonal to the $\xi$-axis. Again, it will be assumed that the reflector normal is fixed in an inertia frame of reference, so that the radiation pressure force will remain directed along the $\xi$-axis, for example through passive spin stabilization, and there are no gradient torques acting on the reflector. An infinitesimal radial displacement $\delta \xi$ is also applied, as in Section 6.4. Then, for a reflector at some equilibrium distance $\bar{\xi}$ with lightness number $\bar{\beta}_D = e^{\mu \bar{\xi}}$, the radial distance from the central star to the reflector is now defined by $r^2 = (\bar{\xi} + \delta \xi)^2 + \delta \eta^2$. From equation (42) it can then be shown that equations of motion for some general position $(\xi, \eta)$ are given by

$$\ddot{\xi} = -\Omega^2 \frac{1}{r^2}\left(\frac{\xi}{r}\right) + \bar{\beta}_D \Omega^2 \frac{\cos^2 \gamma}{r^2} e^{-\mu r} \tag{C1a}$$

$$\ddot{\eta} = -\Omega^2 \frac{1}{r^2}\left(\frac{\eta}{r}\right) \tag{C1b}$$

where $(\xi/r)$ and $(\eta/r)$ define the components of the gravitational acceleration along the $\xi$ and $\eta$ axes. The reduction in radiation pressure exerted on the reflector is defined by $\cos^2 \gamma$ where $\gamma$ is the angle between the radial direction form the central star and the reflector normal, such that $\cos \gamma = (\xi/r)$. Then, for infinitesimal displacements $\xi = \bar{\xi} + \delta \xi$ and $\eta = \delta \eta$, it can be seen that

$$\delta \ddot{\xi} = -\frac{\Omega^2 (\bar{\xi} + \delta \xi)}{\left((\bar{\xi} + \delta \xi)^2 + \delta \eta^2\right)^{3/2}}$$
$$+ \bar{\beta}_D \Omega^2 \frac{(\bar{\xi} + \delta \xi)^2}{\left((\bar{\xi} + \delta \xi)^2 + \delta \eta^2\right)^2} e^{-\mu \sqrt{(\bar{\xi} + \delta \xi)^2 + \delta \eta^2}} \tag{C2a}$$

$$\delta \ddot{\eta} = -\frac{\Omega^2 \delta \eta}{\left((\bar{\xi} + \delta \xi)^2 + \delta \eta^2\right)^{3/2}} \tag{C2b}$$

Expanding equations (C2) to linear order it can then be shown that

$$\delta \ddot{\xi} + \mu \frac{\Omega^2}{\bar{\xi}^2} \delta \xi = 0 \tag{C3a}$$

$$\delta \ddot{\eta} + \frac{\Omega^2}{\bar{\xi}^3} \delta \eta = 0 \tag{C3b}$$

Therefore, both the vertical and transverse dynamics of the problem are found to linearly stable with oscillation frequency $\Omega$, where $\Omega = \sqrt{GM_*/R^3}$. Again, the stability of the equilibria and the decoupling of the radial and transverse dynamics is to be expected. If the position of the reflector is displaced by some small angle $\delta \gamma$ away from the radial direction as seen from the central star, then the radiation pressure force will scale as $f_R \cos^2 \delta \gamma \approx f_R$ if $\delta \gamma \ll 1$. Similarly, the gravitational force will have a radial component $-f_G \cos \delta \gamma \approx -f_G$ and a transverse component $-f_G \sin \delta \gamma \approx -f_G \delta \gamma$. It can be seen that the radial forces remain in equilibrium, which has been demonstrated to be stable, while there is a stable transverse restoring force.

This paper has been typeset from a Microsoft Word file prepared by the author.